\documentclass[manuscript]{aa}

\usepackage{natbib}

\newcommand{\bepo}{{\em BeppoSAX\ }}
\newcommand{\rxte}{{\em RXTE\ }}

\newcommand{\sax}{{ SAX\,J2103.5+4545}}

\newcommand{\swift}{{\em Swift}}
\newcommand{\halfa}{H$_{\alpha}$}

\usepackage{graphicx}
\usepackage{amsmath}
\usepackage{amssymb}

\title{Recent activity of the Be/X-ray binary system SAX\,J2103.5+4545}
\author{A. Camero
     \inst{1}
    \and
    C. Zurita
    \inst{2,3}
    \and
    J. Guti\'{e}rrez--Soto     
    \inst{4,5}
    \and
    M. \"{O}zbey Arabac{\i}
    \inst{6}
    \and
    E. Nespoli
    \inst{7}
    \and 
     F. Kiaeerad
    \inst{8,9}
    \and
     E. Beklen
      \inst{10}
      \and \\
       J. Garc\'{i}a--Rojas
      \inst{2,3}
     \and
     M. Caballero-Garc\'{i}a
     \inst{11}
     }
     \institute{$^1$  Institut de Ci\`{e}ncies de l'Espai, (IEEC-CSIC), Campus UAB, Fac. de Ci\`{e}ncies, Torre C5 pa., 08193, Barcelona, Spain.\\
$^2$ Instituto de Astrof\'{\i}sica de Canarias, E-38200, La Laguna, Tenerife, Spain.  \\
$^3$ Universidad de La Laguna, Dept. Astrof\'{i}sica, E-38206, La laguna, Tenerife, Spain.\\
$^4$ Universitat Internacional Valenciana-VIU, C/Jos\'{e} Pradas Gall\'{e}n s/n, 2$^a$ p., 12006 Castell\'{o}n de la Plana, Spain.\\
$^5$ Instituto de Astrof\'{i}sica de Andaluc\'{i}a (CSIC), Glorieta de la Astronom\'{i}a s/n, 18008, Granada, Spain.\\
$^6$ Department of Physics, Middle East Technical University, Ankara, 06531, Turkey.\\
$^7$ Observatorio Astron\'{o}mico de la Univ. de Valencia, C/Catedr\'{a}tico Jose Beltran, 2, 46980 Paterna (Valencia), Spain. \\
$^8$ Nordic Optical Telescope, Apartado 474, 38700 Santa Cruz de La Palma, Spain.\\
$^9$ Department of Astronomy, Oscar Klein Center, Stockholm University, AlbaNova, Stockholm SE-10691, Sweden.\\
$^{10}$ Physics Department, S\"{u}leyman Demirel University, 32260 Isparta, Turkey.\\
$^{11}$ Czech Technical University in Prague, Faculty of Electrical Engineering, Prague, Czech Republic.\\
}  

\authorrunning{Camero et al.}
\titlerunning{Recent activity of SAX J2103.5+4545}
\offprints{A. Camero\\  \email{camero.ascension@gmail.com}}
\date{Received ; accepted}

\begin{document}

\abstract{}
{We present a multiwavelength study of the Be/X-ray binary system SAX\,J2103.5+4545 with the goal of better characterizing the transient behaviour of this source.  
}
{SAX\,J2103.5+4545 was observed by \textit{Swift}-XRT four times in 2007 from April 25 to May 5, and during quiescence in 2012 August 31. In addition, this source has been monitored from the ground-based astronomical observatories of  El Teide (Tenerife, Spain), Roque de los Muchachos (La Palma, Spain)  and Sierra Nevada (Granada, Spain) since 2011 August, and from the T\"{U}B\.{I}TAK National Observatory (Antalya, Turkey) since 2009 June.  We have performed spectral and photometric temporal analyses in order to investigate the different states exhibited by SAX\,J2103.5+4545. 
}
{In X-rays, an absorbed power law model provided the best fit for all the XRT spectra. An iron-line feature at $\sim$6.42 keV was present in all the observations except for that taken during quiescence in 2012. The photon indexes are consistent with previous studies of SAX\,J2103.5+4545 in high/low luminosity states. Pulsations were found in all the XRT data from 2007 (2.839(2) mHz; MJD 54222.02), but not during quiescence. Both optical outbursts in 2010 and 2012 lasted for about 8/9 months (as the one in 2007 probably did and the current one in 2014 might do) and were most probably caused by mass ejection events from the Be star that eventually fed the circumstellar disc. All of these outbursts started about 3 months before the triggering of the X-ray activity, and about the same period before the maximum of the H$\alpha$ line equivalent width (in emission) was reached at only $\sim$ -5\,\AA.  In this work we found that the global correlation between the BV variability and the X-ray intensity was also observed at longer wavelengths in the IR domain.}
{}
{}

\maketitle

\section{Introduction}

%%%%%   LOG OBSERVATIONS ######
\begin{figure}
\hspace*{-0.5cm}\includegraphics[width=9.8cm,height=6.7cm]{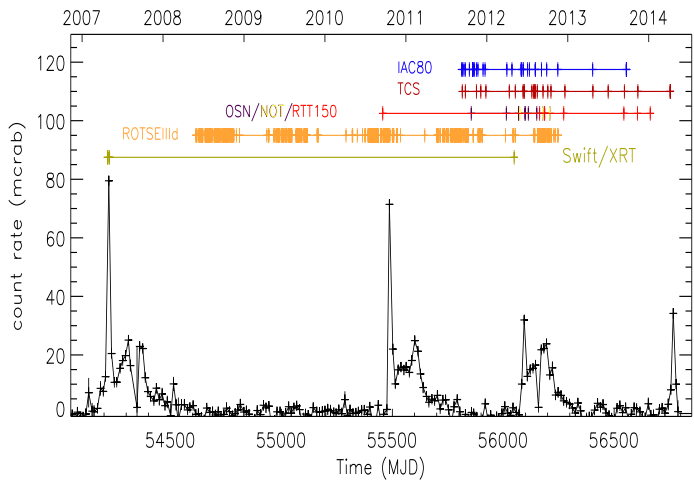}
\caption{\textit{Swift}/BAT light-curve (15--50 keV) with a bin size equal to the 12.67\,d orbital period.  Tick marks on the segments located above the light-curve  denote the times of the \textit{Swift}-XRT pointing observations (green), as well as the optical/IR photometric data from the ground-based telescopes  ROTSEIIId (orange), IAC80 (blue) and TCS (red). The optical spectroscopic observations come from OSN (purple), NOT (yellow) and RTT150 (light red) (see also Table~\ref{ew_tab}).}
\label{plot_log}
\end{figure}
%%%%%%%%%%%%%%%%%%%%%%%%

Accreting X-ray pulsars are binary systems composed of a donor star and an accreting neutron star.  In High Mass X-Ray Binary (HMXB)  systems the optical companion could be either a massive early-type supergiant (supergiant systems) or an O,B main sequence or giant star (BeX binaries; BeXB). Among the most remarkable signatures found in BeXBs are the detection of IR excess and  emission-line features in their optical spectra produced in a disc-like outflow around the Be star.  Historically, their outbursts have been divided into two classes. Type I (or normal) outbursts  normally peak at or close to periastron passage of the neutron star  (L$_X\leq$ 10$^{37}$\,erg s$^{-1}$).  Type II (or giant) outbursts reach luminosities of the order of the Eddington luminosity  (L$_X\sim$10$^{38}$\,erg s$^{-1}$) \citep{frank02}, with no preferred orbital phase. 

The Be/X-ray binary SAX\,J2103.5+4545 was discovered with the \bepo satellite \citep{Hulleman98}, showing X-ray pulsations of $\sim$358\,s. The X-ray spectrum (2--25 keV) was fitted by an absorbed power-law with  a photon index of 1.27$\pm$0.14 (N$_H$=3.1$\times10^{22}$ cm$^{-2}$).  This source has one of the shortest orbital known BeXB periods ($\sim$12.68\,d)  \citep{baykal00}. Together with  1A\,1118-616 \citep{staubert11}, it is the only other system of its type to occupy the region of the wind-fed supergiant binaries in the  P$_{spin}$--P$_{orb}$ diagram. 

The  optical  counterpart is a B0Ve star (V = 14.2) at a distance between 3.2 and 6.5 kpc \citep{baykal02,reig04}. Baykal et al. (2002) found  a correlation between spin-up rate and X-ray flux during the 1999 outburst. This suggested the formation of an accretion disc during periastron passage of the neutron star. Later on, using \textit{XMM} observations a quasi-periodic  oscillation (QPO) at 22.7 s was discovered by \.{I}nam et al. (2004).  

SAX\,J2103.5+4545 shows periodic type I outbursts and stronger events that, despite barely reaching luminosities of the order of 10$^{37}$erg s$^{-1}$, may be called type II outbursts.  A recurrent characteristic of this source has been the detection of a flare as a precursor of a type II outburst.  Two years after its discovery, \sax~was found to be active again in 1999 October by the All-Sky Monitor on board \rxte \citep{baykal00}.  A large flare followed by the main outburst was observed by \rxte in 2001 March \citep{baykal02} and in 2002 June \citep{baykal07, camero07}.  

\textit{INTEGRAL} started observing this source in 2002 December \citep{lutovinov03, filippova04, blay04, sidoli05, falanga05,camero07, ducci08, reig10}. After three years in a low luminosity state, the source was detected in 2007 April during \textit{INTEGRAL} observations of the Galactic plane \citep{galis07}, anticipating a large outburst also observed by \swift~ \citep{krimm07, kiziloglu09}.  Coinciding with the renewed X-ray activity,  the optical counterpart showed an H$_{\alpha}$ line in emission in 2007 May, evidence of disc formation around the Be star \citep{Manousakis07, kiziloglu09, reig10}. 

A re-brightening of the Be star in 2010 August indicated an imminent outburst in X-rays \citep{kiziloglu10}. The \textit{Swift}/BAT monitor confirmed a new X-ray episode starting in 2010 October \citep{Krimm10}.  Renewed optical activity  was reported in 2012 April (Konstantinova $\&$ Blinov 2012). Two months later, the system was detected by \textit{INTEGRAL} \citep{sguera12}. In addition, contemporaneous observations in the JHKs filters showed an IR enhancement, and the H$_{\alpha}$ line was found in emission (Camero-Arranz et al. 2012).  Recently, renewed optical activity of this system has been reported in 2014 March (Konstantinova $\&$ Mokrushina 2014). Since April, X-ray activity is also being detected for this system by several X-ray missions. SAX\,J2103.5+4545 continues in outburst at the moment of writing the present work.

Here, we present a multiwavelength study  of \sax~since 2010. For this purpose we used optical/IR data from our dedicated campaign involving several ground-based astronomical observatories. In addition, we used one \textit{Swift}-XRT pointed observations carried out in 2007 and 2012, and survey data from different space-borne telescopes.

 %%%%%%%%  REFERENCE STARS  %%%%%%
\begin{table}
\caption{Selected photometric reference stars in the neighbourhood of the optical counterpart of SAX J2103.5+4545 (GSC\,03588-00834) with the ROTSEIIId telescope.}                          
\begin{tabular}{lccc} 
\hline\hline 
\hspace{-0.2cm} Star &\hspace{-0.4cm}  RA   &\hspace{-0.4cm}  Dec  &\hspace{-0.3cm}	\textsc{USNO-A2}  \\
        star		              &\hspace{-0.4cm}  (J2000.0) &\hspace{-0.4cm}   (J2000.0)  & 	 R mag \\
\hline\hline
GSC\,03588-00834     &    21$^h$03$^m$35$^s$.7          & \hspace{-0.3cm} +45$^{\circ}$45$'$04$"$.0           &    14.4     \\
Star 1  &   21$^h$03$^m$20$^s$.58  &\hspace{-0.3cm} +45$^{\circ}$43$'$09$"$.1  &	 12.9 \\
Star 2  &  21$^h$03$^m$21$^s$.99  &\hspace{-0.3cm}+45$^{\circ}$44$'$55$"$.5  &	13.8\\
Star 3  & 21$^h$03$^m$40$^s$.32  &\hspace{-0.3cm} +45$^{\circ}$46$'$22$"$.6  &	13.0\\
\hline \smallskip                        
\label{compar}  
\end{tabular}
\end{table}
%%%%%%%%%%

  %%%%%%%%  EW 
\begin{table*}
\caption{H$_{\alpha}$ line equivalent width (EW) measurements for SAX J2103.5+4545. Negative values indicate that the line is in emission.}                          
\begin{center}
\begin{tabular}{lcccccc} 
\hline\hline
DATE      &  MJD       & Telescope$^a$ & EW  & FWHM & Profile$^b$  &  $\Delta$V$^c$\\
&       &     &  (\AA)  &  (\AA) &  & $\pm$50)  \\
\hline \hline

2010-Sep-17 &55456.707 &RTT150 &-1.02$\pm$0.08 & 9.47$\pm$0.10 &V$\approx$R&365.63 \\
2011-Oct-22 &55856.007 &OSN    &+1.44$\pm$0.12 & 8.45$\pm$0.26 &ABS	   &--      \\
2012-Mar-28 &56014.133 &OSN    &-1.34$\pm$0.24 &10.36$\pm$0.32 &V$<$R	   &494.74  \\
2012-May-22 &56069.168 &OSN    &-1.73$\pm$0.21 &15.94$\pm$0.31 &V$>$R	   &498.86  \\
2012-May-23 &56070.170 &OSN    &-1.07$\pm$0.26 &15.26$\pm$0.08 &V$\approx$R&445.33  \\
2012-May-28 &56075.101 &NOT    &-1.19$\pm$0.28 &17.07$\pm$0.14 &V$>$R	   &429.81  \\
2012-Jun-20 &56098.058 &OSN    &-1.57$\pm$0.30 &14.04$\pm$0.15 &V$>$R	   &363.29  \\
2012-Jun-20 &56098.996 &OSN    &-1.88$\pm$0.35 &14.13$\pm$0.36 &V$<$R	   &394.94  \\
2012-Jul-05 &56113.058 &OSN    &-0.94$\pm$0.18 &11.84$\pm$0.36 &V$>$R	   &371.55  \\
2012-Jul-29 &56137.180 &NOT    &-1.41$\pm$0.19 &11.47$\pm$0.30 &V$<$R	   &334.22  \\
2012-Aug-12 &56151.096 & OSN   &-2.19$\pm$0.18 &11.98$\pm$0.38 &V$>$R	   &301.07  \\
2012-Aug-24 &56163.952 &RTT150 &-3.00$\pm$0.06 &10.28$\pm$0.24 &V$>$R	   &283.86  \\
2012-Sep-08 &56178.071 &OSN    &-4.44$\pm$0.26 &18.52$\pm$0.52 &SPE	   &--      \\
2012-Sep-09 &56179.134 &NOT    &-3.85$\pm$0.24 &17.71$\pm$0.35 &V$>$R	   &442.71  \\
2012-Sep-15 &56185.919 &RTT150 &-3.87$\pm$0.16 &10.00$\pm$0.08 &V$>$R	   &383.86  \\
2012-Sep-16 &56186.898 &RTT150 &-4.56$\pm$0.09 &14.09$\pm$0.37 &V$>$R	   &387.64  \\
2012-Oct-10 &56210.017 &OSN    &-4.37$\pm$0.24 &12.03$\pm$0.32 &V$>$R	   &390.89  \\
2012-Dec-11 &56272.703 &RTT150 &-0.70$\pm$0.12 & 4.73$\pm$0.46 &V$<$R	   &209.51  \\
2013-Sep-08 &56543.791 &RTT150 &+1.91$\pm$0.68 &10.66$\pm$0.69 &ABS	   &--      \\
2013-Nov-08 &56604.870 &RTT150 &+2.26$\pm$0.40 &11.81$\pm$0.26 &ABS	   &--      \\
2014-Jan-05 &56662.733 &RTT150 &+0.96$\pm$0.14 & 5.75$\pm$0.41 &ABS	   &--      \\
\hline                       
\label{ew_tab}  
\end{tabular}

\hspace*{+1cm}$^a$Resolution of the spectra for each instrument. RTT150: R$\approx$2200; OSN: R$\approx$1400; NOT: R$\approx$1300--2600.\\
\hspace*{-6.9cm}$^b$ABS, absorption; SPE, single-peaked emission.\\
\hspace*{+1.35cm}$^c\Delta$V, separation between the violet and red peaks of the double-peaked emission profiles. Units in km\,s$^{-1}$.\\
\end{center}
\end{table*}
%%%%%%%

\section{Observations and  Data Reduction}\label{observations}
 
\subsection{Optical/Infrared Observations}

SAX J2103.5+4545 has been continually monitored in the optical and infrared during the period 2011 September to 2014 May with the 80-cm IAC80 and with the 1.5-m TCS telescopes at the Observatorio del Teide on Tenerife (Spain). In the optical, we obtained CCD images in the B and V bands with integration times of 30\,s. In infrared, J, H and K-short simultaneous observations were performed using the CAIN camera with integration times of 150\,s. All images were reduced in the standard way using the pipelines available for both telescopes. We applied straightforward aperture photometry using apertures of 1.5 times the full width at half maximum (FWHM). Several comparison stars within the field of view were checked for variability during each night and throughout the entire data set. Calibration of the optical data was performed using the zero point magnitude offsets and extinction coefficients listed for the Observatorio del Teide. Infrared data were calibrated using the 2MASS survey as a photometric reference. Figure~\ref{plot_log} summarizes the time span of these observations (see also Tables \ref{opt_phot} and \ref{IR_phot}). 

In addition, photometric data covering 2009 June 18 to 2012 November 15 were obtained with the 0.45-m ROTSEIIId\footnote{The Robotic Optical Transient Search Experiment, ROTSE, is a collaboration of Lawrence Livermore National Lab, Los Alamos National Lab, and the University of Michigan (http://www.rotse.net)} telescope at the T\"{U}B\.{I}TAK National Observatory (Antalya, Turkey), which operates without filters with a wide pass-band peaking at 5500 \AA~ \citep{akerlof03}. Reference stars located in the surroundings of the optical counterpart of the system, CSC\,03588--00834, were selected showing relatively constant magnitudes (see Table~\ref{compar}) in the ROTSEIIId pass-band. In order to obtain the differential magnitudes for the Be star we subtracted the reference star's magnitude from the Be star's magnitudes (see Sect.~\ref{OIR}). More details on the reduction of the ROTSEIIId data can be found in \cite{kiziloglu05}. 

Optical spectroscopic observations, covering the period from 2010 August to 2014 January (see Fig.~\ref{plot_log} and Table~\ref{ew_tab}), were obtained from three different telescopes: the
2.56-m Nordic Optical Telescope (NOT) located at the Observatorio del Roque de los Muchacho (La Palma, Spain), the 1.5-m Telescope at the Observatorio de Sierra Nevada (OSN-CSIC) in Granada (Spain), and the Russian-Turkish 1.5-m telescope (RTT150) at the T\"{U}B\.{I}TAK National Observatory in Antalya (Turkey). %\textbf{We also include recent spectroscopical observations from 2014 May from theSpanish Astronomical Observatory of the University of Valencia (OAO).}

The spectra from NOT were obtained with the Andaluc\'ia Faint Object Spectrograph and Camera ALFOSC,\footnote{The data presented here were obtained [in part] with ALFOSC, which is provided by the Instituto de Astrof\'isica de Andaluc\'ia (IAA) under a joint agreement with the University of Copenhagen and NOTSA.}  using Grism\,7, with a dispersion of 1.5\,\AA/pixel, and 0.5"--1" slits, whereas the low resolution OSN spectra were acquired using Albireo spectrograph. The Albireo spectra have a resolution of  R$\approx$1400 and are centred on the H$_{\alpha}$. %
The reduction of this data set was performed  using standard procedures within IRAF,\footnote{IRAF is distributed by the National Optical Astronomy Observatory, optical images which is operated by the Association of Universities for Research in Astronomy (AURA) under cooperative agreement with the National Science Foundation.} including bias subtraction, removal of pixel-to-pixel sensitivity variations, optimal spectral extraction, and wavelength calibration based on arc-lamp spectra. Finally, the spectroscopic data from  RTT150 were obtained with the T\"{U}B\.{I}TAK Faint Object Spectrometer
(TFOSC) equipped with a 2048$\times$2048, 15$\mu$m pixel Fairchild 447BI CCD. We used
Grism 8 having an average dispersion of 1.4 \AA/pixel and 67$\mu$m slit (1.24") providing
a 5800--8300 \AA wavelength coverage. The reduction of RTT150 spectra was done using the Long-Slit package of MIDAS.\footnote{http://www.eso.org/projects/esomidas}  The corrections of bias and 
flat-field and removal of cosmic-ray hits were carried out with standard MIDAS routines. All spectroscopic data
were normalized with a spline fit to continuum and corrected to the heliocentre after
the wavelength calibration. The full width at H$_\alpha$ maximum (FWHM) and equivalent width (EW) measurements were acquired by fitting Gaussian functions to the H$_\alpha$ profiles using the ALICE subroutine of MIDAS.

\subsection{X-ray Observations}\label{obs}

\sax\, was observed by \textit{Swift}-XRT \citep{gehrels04,burrows05} four times from  2007 April 25 (MJD 54215.4) to 2007 May 5 (MJD 54225.1), and once again in 2012 August 31 (MJD 56170.4) for a total exposure time of 8.56\,ks and  3.13\,ks, respectively.  Some of these observations were performed in Photon Counting (PC) mode (2.5\,s time resolution, 2D Image) and/or in Windowing Timing (WT) mode (1.7 ms time resolution, 1D Image). Figure~\ref{plot_log} and Table~\ref{log_xrt} provide the log of these observations.

For all the observations screened event files were created with the \texttt{xrtpipeline} tool (HEAsoft version 6.13) and  the events converted to the solar system barycentre. Only XRT data in PC mode suffered from pile-up (see Table~\ref{log_xrt}). We  thus extracted source events from an annulus circle region of 30 pixels outer radius (1 pixel = 2.36$"$) and 3 pixels inner radius.  The background was extracted selecting a larger region of $\sim$60 pixel radius but far from the source. For the WT observations we extracted source events from a box (30$\times$15 pixel) region of the one-dimensional image, centred on the pixel with the highest rate. The background was chosen using two segments on both sides of the source, but far enough not to get contaminated by source photons. The spectra were then generated  using \texttt{xselect}.   The ancillary response matrix was created using the \texttt{xrtmkarf} task.  The relevant response matrix to use is given by the HEASARC's calibration database (CALDB\footnote{http://heasarc.gsfc.nasa.gov/FTP/caldb}).  We also used for comparison data products supplied by the UK \textit{Swift} Science Data Centre at the University of Leicester \citep{evans09}.

Since 2008, the Gamma-ray Burst Monitor (GBM) on board the \textit{Fermi} satellite, has been monitoring \sax. In this study we  used timing products provided by the GBM Pulsar Team \citep[see e.g.][for a detailed description of the timing technique]{finger09, camero10}.  We also used quick-look X-ray results provided  by the \textit{RXTE}  All Sky Monitor team,\footnote{http://xte.mit.edu/ASM$\_$lc.html}  and \textit{Swift}/BAT transient monitor results provided by the \textit{Swift}/BAT team \citep{krimm13}.

\section{Results}

%%%%%  OPTICAL/IR
\begin{figure}
\hspace*{-0.4cm}\includegraphics[width=9.5cm,height=11.25cm]{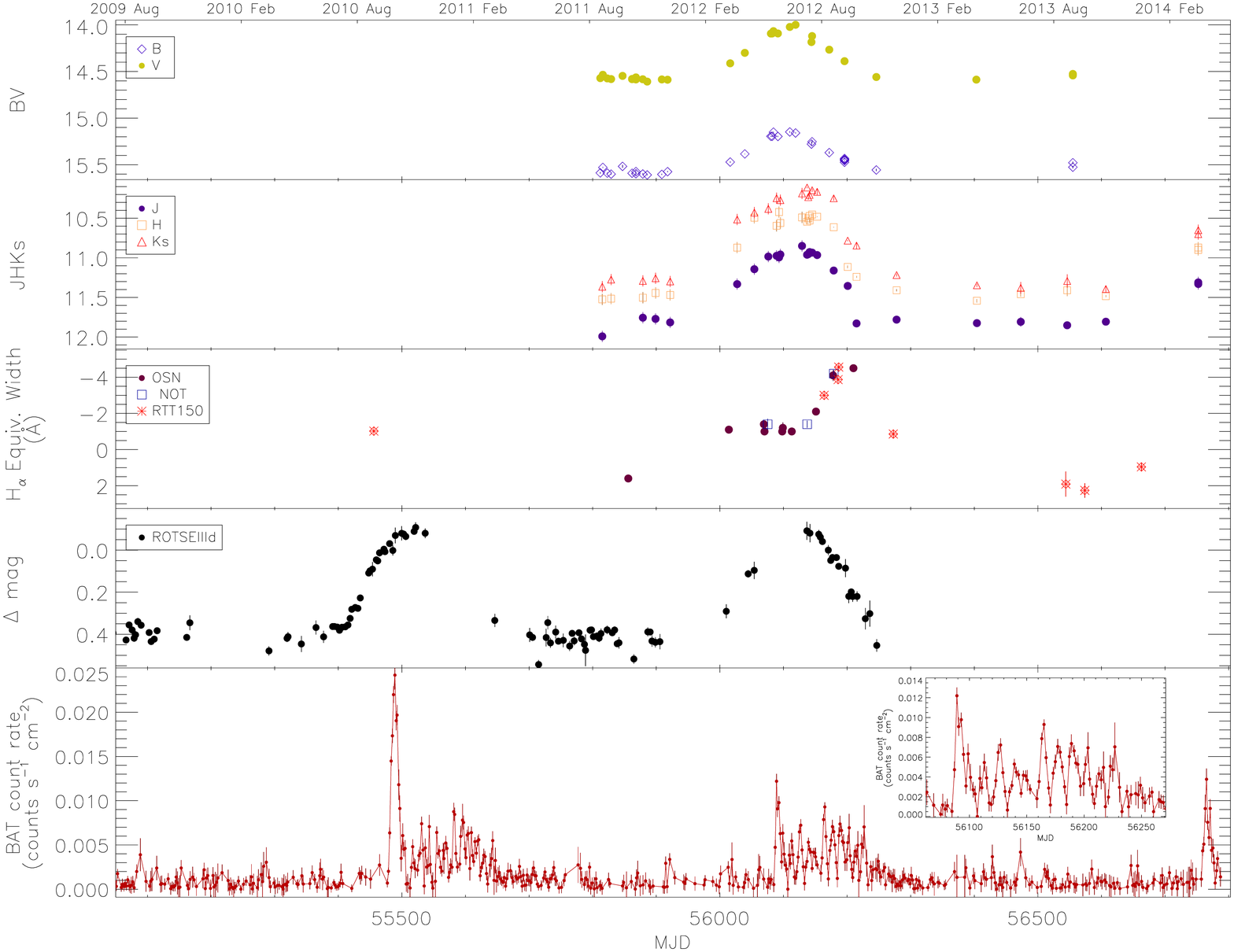}
\caption{Optical/IR and X-ray evolution of SAX\,J2103.5+4545. The time bin for the BAT lightcurve is 2 days. There is a notable correlation between the variability in the optical/IR and X-ray bands.The correlation of the X-ray flux with the ~12.7 orbital period (only during bright state) can be better appreciated in the small panel displaying a zoom of the 2012 X-ray outburst.}
\label{opt_ir}
\end{figure}
%%%%%%%%%%%%%%%%%%%%%%%%

\subsection{Optical/IR  Photometry}\label{OIR}

%%%%%%%%  XRT LOG OBS  %%%%%%
\begin{table}
\caption{Log of the \textit{Swift}-XRT observations.}                          
\begin{tabular}{llllllll} 
\hline\hline 
\hspace{-0.2cm}ObsID &\hspace{-0.2cm}  MJD  	&\hspace{-0.3cm}  Exp.$^{a}$ 	&\hspace{-0.32cm}  Exp.$^{a}$  &\hspace{-0.32cm}	  Exp.$^{a}$  &\hspace{-0.35cm}	 Rate 	& $\nu_{spin}$ & PF$^b$	\\
\hspace{-0.35cm} \textsc{(30922)} &\hspace{-0.29cm}    &\hspace{-0.3cm}  Total         &\hspace{-0.25cm}  PC 	 &\hspace{-0.32cm}	  WT &\hspace{-0.33cm}    (c/s) & \hspace{-0.2cm} (mHz) & ($\%$ ) \\
\hline \hline 
\hspace{-0.05cm}001  &\hspace{-0.55cm}54215.44  &\hspace{-0.25cm} 1.65  &\hspace{-0.33cm} 0.25 &\hspace{-0.4cm} 1.40  &\hspace{-0.4cm} 6.17(7)  &\hspace{-0.39cm} 2.834(22) &\hspace{-0.39cm} 33(4)\\
\hspace{-0.05cm}002  &\hspace{-0.55cm}54216.45  &\hspace{-0.25cm} 1.54  &\hspace{-0.33cm} 0    &\hspace{-0.4cm} 1.54  &\hspace{-0.4cm} 9.54(8)  &\hspace{-0.39cm} 2.835(11) &\hspace{-0.39cm} 41(6)  \\
\hspace{-0.05cm}003  &\hspace{-0.55cm}54222.02  &\hspace{-0.25cm} 2.54  &\hspace{-0.33cm} 2.53 &\hspace{-0.4cm} 0.01  &\hspace{-0.4cm} 0.81(2)  &\hspace{-0.39cm} 2.839(2)  &\hspace{-0.39cm} 37(5) \\
\hspace{-0.05cm}004  &\hspace{-0.55cm}54225.09  &\hspace{-0.25cm} 2.82  &\hspace{-0.33cm} 2.55 &\hspace{-0.4cm}	0.27  &\hspace{-0.4cm} 0.55(2)  &\hspace{-0.39cm} 2.832(5)  &\hspace{-0.39cm} 31(6) \\
\hspace{-0.05cm}005  &\hspace{-0.55cm}56049.09  &\hspace{-0.25cm} 3.13  &\hspace{-0.33cm} 0    &\hspace{-0.4cm} 3.13  &\hspace{-0.4cm} 0.074(9) &\hspace{-0.3cm}     --	     &\hspace{-0.39cm} 64(35)  \\

\hline                        
\end{tabular}

$^{a}$ units in ks.\\
$^b$Pulse fraction (PF) defined as F$_{max}$--F$_{min}$/F$_{max}$+F$_{min}$ (further details in Sect.~\ref{timing}).\\
\label{log_xrt}  
\end{table}
%%%%%%%%%%

The results of our multiwavelength campaign are shown in Figure~\ref{opt_ir}. In this figure the optical/IR behaviour of the Be star is shown, together with the X-ray activity observed from the neutron star (see Tables \ref{opt_phot} and \ref{IR_phot}). A correlation between the variability in the optical/IR bands and the X-ray intensity is clearly noticeable.

From top to bottom, the fourth panel of the same figure shows a steady increase in the ROTSEIIId optical differential magnitude from 2010 May to August (MJD $\sim$55320--55410). The rise in the brightness of the Be star indicated the beginning of the optical outburst, which approximately peaked on 2010 November 20 (MJD $\sim$55520) and most probably ended around 2011 April. About three months after the optical reawakening, the X-ray activity was triggered with a large flare followed by periodic 12.67\,d X-ray outbursts overlapping the increased X-ray emission for about 6--7 months (see bottom panel). 

According to the evolution of the BVJHKs magnitudes (see the first and second panels of Fig.~\ref{opt_ir}), SAX\,J2103.5+4545 exhibited a steady trend from 2011 November to 2012 March (MJD $\sim$55835--56000).  Both optical and IR brightenings then took place, peaking around 2012 July (MJD 56130) and finished by November (MJD$\sim$56245). At that point, the system went back to its quiescent state (see fourth panel of the same figure). About three months after the triggering of optical/IR activity a large X-ray flare took place in 2012 May/June (MJD $\sim$56070), followed by the same behaviour pattern previously observed in 2010 and 2007.  The compact object went back to quiescence $\sim$7 months later. Renewed optical/IR activity of this system started most likely in 2014 February. About 2--3 months later X-ray activity was detected for this system by \textit{Swift}/BAT\footnote{http://swift.gsfc.nasa.gov/results/transients} and \textit{Fermi}/GBM\footnote{http://gammaray.nsstc.nasa.gov/gbm/science/pulsars}.

Therefore, both optical outbursts in 2010 and 2012 lasted for about 8/9 months, and the Be star reached maximum brightness after X-ray maximum.  At this moment, the current behaviour of this source resembles so far the activity displayed in 2012. As in K{\i}z{\i}lo{\v g}lu et al. (2009),  we searched for periodicity in the ROTSEIIId light-curve, but nothing was observed on shorter time scales. All the available data for this source in this band show that the recurrence of the optical outbursts is highly variable. The time span is $\sim$5  to $\sim$3 yr (see also Reig et al. 2010, K{\i}z{\i}lo{\v g}lu et al. 2009),  or, more recently,  less than 2 yr.

On average, the brightening of the system in the IR was  about 1.4 magnitudes. In the optical band the amplitude was $\sim$0.6 magnitudes for the 5500 \AA~band-pass, and 0.4 magnitudes for the BV filters. This confirms the increasing trend in the amplitude of this variation with wavelength, interpreted as a contribution from the Be star's disc \citep[and references therein]{reig10}. 

\subsection{H$_{\alpha}$ Line} \label{halpha}

\begin{figure}
\hspace{-1cm}\rotatebox{-90}{\includegraphics[width=8cm,height=10.6cm]{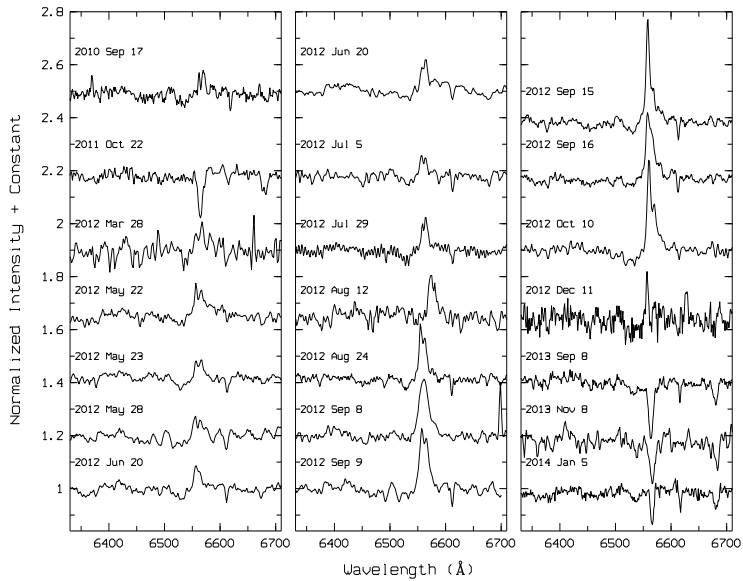}}
\caption{ H$_{\alpha}$ line profile evolution for SAX\,J2103.5+454.  The intensity of each profile was normalized and scaled by adding a constant. }

\label{ew_evol}
\end{figure}

The emission lines in the spectra of Be stars arise from recombination in a circumstellar disc, while absorption profiles originate in the photosphere.
The exact date on which the H$_{\alpha}$ line changed from being in absorption to being in emission (see third panel of Fig.~\ref{opt_ir}, Fig.~\ref{ew_evol}, and Table~\ref{ew_tab}) is not clear from our data. However, from our data for 2010 September and 2012 March we found that a circumstellar disc around the Be star was already present. In 2012, the \textit{INTEGRAL} satellite reported renewed X-ray activity from this source around two months after the detection of the H$_{\alpha}$ line in emission \citep[][and references therein]{camero12}.

%%%%%  Halpha
%%%%%%%%%%%%%%%%%%%%%%%%

In 2012, the H$_{\alpha}$ EW showed a sudden rise at the time of the X-ray flare and remained practically constant for about one month. The EW increased afterwards, finally peaking in 2012 October. In contrast, the available spectroscopic data from 2007 \citep{reig10} show that the H$_{\alpha}$ EW did not remained flat during the X-ray flare, but showed an initial peak before reaching maximum in 2007 September. Therefore, the amount of mass ready to be transferred from the disc to the neutron star was most probably lower during the 2012 X-ray flare, and so was the flare. Our next measurement of the EW  in 2012 December (MJD 56272.7) confirmed a dramatic decrease in the size of the Be disc, recovering a similar value to the initial steady period.  The H$_{\alpha}$ line has been found in absorption since 2013 September up to recently. %\textbf{In 2014 May the H$_{\alpha}$ was again detected in emission, coinciding with the reawakening of this system sometime around 2014 February.}  

In Figure~\ref{ew_evol} we show the evolution in the shape of this line. We see only absorption profiles at the beginning and the end of the optical/IR activity, an indication that no disc was present. The shape of the H$_{\alpha}$ line is highly variable with some emission profiles showing asymmetric single- and double-peaked shapes. In addition, double-peaked profiles merged into a single peak as the EW/luminosity increase.

\begin{figure}
\begin{center}
\hspace{-0.2cm}\includegraphics[width=8cm,height=5cm]{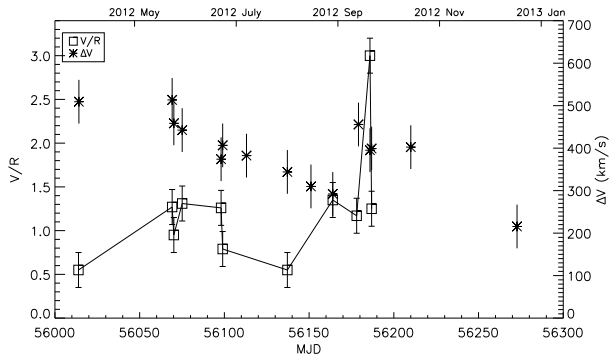}
\caption{V/R ratios (open squares) of the H$\alpha$ line during the most recent period of activity in 2012. Rapid changes in the temporal V/R evolution can be observed. The peak separation $\Delta$V (stars) is also overplotted.}
\label{VR}
\end{center}
\end{figure}

For the typical double-peaked emission lines, the heights of the violet and redshifted peaks are designated as V and R, respectively \citep{porter03}. Figure~\ref{VR} shows the V/R ratios of the H$\alpha$ line for SAX J2103.5+4545 since 2010 September (see also Table~\ref{ew_tab}). Although the coverage of the data is non-uniform, the former figure reveals very rapid changes on short times scales. The most remarkable change occurred towards the end of the growing cycle of the disc. Around 2012 September 15, the V/R ratio changed by a factor of $\sim$2.5 within a day. 

We also measured the distance between the V and R peaks - that is, $\Delta$V - of the H$_\alpha$ emission line by fitting two Gaussian functions to the profiles (see Table~\ref{ew_tab}).  Figure~\ref{VR}  shows the evolution of $\Delta$V during the most recent active period  in 2012. Initially, $\Delta$V decreases as the EW increases exponentially (see also the third panel of Fig.~\ref{opt_ir}).  A sudden increase in $\Delta$V was then observed around 2012 September 9 (MJD 56179), decreasing afterwards. This rapid change took place almost at the same time as the one observed in the V/R ratios. It seems that the Be disc might have been truncated at that time. 

It is known that Be stars are rapid rotators, and that their rotational velocities can be measured either using HeI lines  between 4026--4471 \AA ~\citep{steele99} or H$_\alpha$
lines. We have used H$_\alpha$ measurements to find the projected rotational velocity (v\,sin\,i) of this source, since our data do not cover the blue part of the spectrum. \cite{hanuschik89} gives the relation between the H$_\alpha$ widths and the rotational velocity as follows: log[FWHM(H$_\alpha$)/1.23\,v\,sin\,i + 70] = -0.08\,logEW(H$_\alpha$) + 0.14, where EW and FWHM are in \AA, and v\,sin\,i are in km s$^{-1}$ (see Table~\ref{ew_tab}). Using the average EW and FWHM values of the H$_\alpha$ line, we have calculated the projected rotational velocity of SAX\,J2103.5+4545 as v\,sin\,i$\sim$292 km\,s$^{-1}$, which is of the same order as the value of 240 km\,s$^{-1}$ reported by Reig et al. (2004).

%%%%%  VR variations
%%%%%%%%%%%%%%%%%%%%%%%%

\subsection{ X-ray Activity}\label{xrays}

\subsubsection{Spectral Analysis}

To  obtain the 0.3--10\,keV phase-averaged spectrum for the XRT observation during quiescence in 2012 we used source and background photons extracted as described in Section~\ref{obs}. For the present analysis we used the {\tt XSPEC} package (version 12.8.0j) \citep{arnaud96}. Despite the poor statistics, an absorbed power law (PL) model was found to be the best fit to the quiescent spectrum. For the photoelectric absorption we used the cross-sections from \cite{balucinska92} and the Solar abundance from \cite{anders_grevesse89}. The best-fit parameters for this model are $N_{\rm H} <$0.02$\times10^{22}$ cm$^{-2}$ (this parameter was not  well constrained, being compatible with zero), and $\Gamma$=2.8$\pm$0.8 (C-stat=184.7 for 162 d.o.f.).  Results from data in PC and WT modes were found to be compatible within the uncertainties. The 2--10 keV observed flux was F$_X$= 0.005(2)$\times$10$^{-10}$\,erg cm$^{-2}$ s$^{-1}$.  The rightmost column of Table~\ref{spec_tab} contains the main spectral parameters obtained (ObsID 005).  

\begin{figure}
\hspace{-0.7cm}\rotatebox{-90}{\includegraphics[width=6.5cm,height=9.5cm]{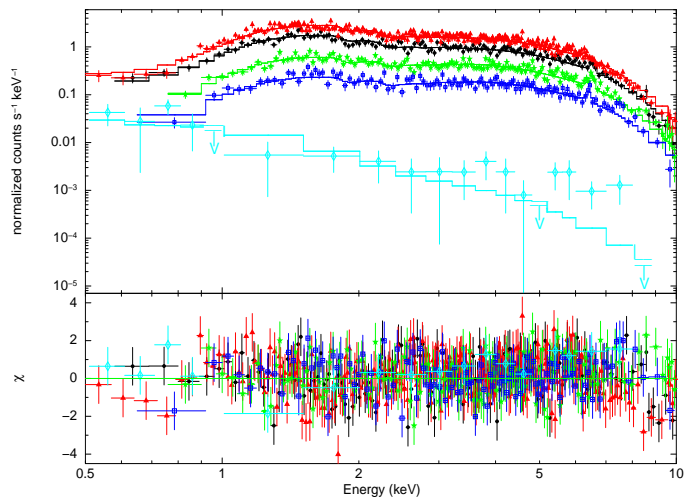}}
\caption{\swift-XRT spectra for all the observations. For clarity, the spectra have been grouped with a minimum of 25 counts per bin. ObsID 001 data poins are denoted as black circles, ObsID 002 as red triangles, ObsID 003 as light green stars, ObsID 004 as dark blue open squares and ObSID 005 as turquoise diamonds.}
\label{spec}
\end{figure}

To study the spectral evolution of SAX\,J2103.5+4545, we fitted the spectra from 2007, again with an absorbed PL model (see  Fig.~\ref{spec}), since this provided the best results. We added an iron-line feature at $\sim$6.42 keV, which had been previously observed in this source (see  Table~\ref{spec_tab}).  We fixed the width of iron line to the obtained value to better constrain the errors in its energy. In the 2--10 keV range the observed flux was F$_X =$10.4(2), 12.2(1), 5.9(1), and 4.9(1)$\times$10$^{-10}$\,erg cm$^{-2}$ s$^{-1}$ for ObsIds 001, 002, 003 and 004, respectively.  The observed flux in the 0.5--10\,keV band dropped from 1.08(2)$\times$10$^{-9}$ erg cm$^{-2}$\,s$^{-1}$ at the maximum of the 2007 flare, to 0.0020(5)$\times$10$^{-9}$ erg cm$^{-2}$\,s$^{-1}$  during the quiescent state in 2012. In addition, we froze the $N_{\rm H}$  during quiescence using average values ranging from 0.5 to 0.8$\times10^{22}$ cm$^{-2}$, resulting in worst fits (e.g. C-stat/d.o.f.=218.3/163 for $N_{\rm H}$ =0.8$\times10^{22}$ cm$^{-2}$) comparing to that in which all the parameters were set free (C-stat/d.o.f.=184.7/162).  We also found a remarkable deviation of the model with respect to the data points at low energies. We would like to point out that the poor quality of this spectrum is probably not only related to the very low flux level of the source  ($\sim$0.074 mCrab), but also to the fact that this short $\sim$3 ks observation was taken in WT mode, more suitable for timing analysis and  for source fluxes between 1--600 mCrab.

The photon indexes ($\Gamma<$1) are consistent with the values previously reported for \sax~ being at high luminosities \citep{inam04, baykal02,baykal07,filippova04, camero07,reig10}.  At low states, our value of the photon index during quiescence is slightly greater than previous measurements, although still compatible taking into account the uncertainties.

%%%%%  SPECTRA
%%%%%%%%%%%%%%%%%%%%%%%%

%%%%%%%%  SPECTRAL FIT  %%%%%%
\begin{table}
\caption{Best spectral parameters from an absorbed PL model plus a Gaussian iron line. Errors are given at the 90$\%$ confidence level.}                          
\begin{tabular}{llllll} 
\hline\hline

\hspace{-0.1cm}Par./Obs. 	 &\hspace{-0.2cm}  001  &\hspace{-0.1cm}  002 	&\hspace{-0.1cm} 003 &\hspace{-0.1cm} 004 &\hspace{-0.1cm} 005\\
\hline \hline 
\hspace{-0.1cm}$N^a_{\rm H}$     &\hspace{-0.4cm}  0.82(3)  &\hspace{-0.4cm}  0.73(2)  &\hspace{-0.4cm} 0.79(5)  &\hspace{-0.4cm} 0.5(3)  &\hspace{-0.4cm}  $<$\,0.02$^*$   \\    %0.8 (fixed) \\
\hspace{-0.1cm}$\Gamma$   	 &\hspace{-0.4cm}  0.86(3)  &\hspace{-0.4cm}  0.89(2)  &\hspace{-0.4cm} 0.85(5)  &\hspace{-0.4cm} 0.6(2)  &\hspace{-0.4cm}  2.8(8)   \\     %1.5(8)	 \\  
\hspace{-0.1cm}$\Gamma_{norm}^b$ &\hspace{-0.4cm}  6.0(1)   &\hspace{-0.4cm}  6.8(2)   &\hspace{-0.4cm} 4.1(2)   &\hspace{-0.4cm} 2.7(8)  &\hspace{-0.4cm}  0.03(1)	 \\
															  
\hspace{-0.25cm} E$_{Fe}$ (keV)  &\hspace{-0.4cm} 6.49${+0.09\atop-0.2}$ &\hspace{-0.3cm} 6.12${+0.3\atop-0.11}$ &\hspace{-0.3cm} 6.48${+0.9\atop-0.18}$  &\hspace{-0.3cm} 6.43${+0.11\atop-0.09}$  &  -    \\
\hspace{-0.15cm}$ \sigma$ (keV)  &\hspace{-0.4cm} 0.1$^c$   &\hspace{-0.4cm}  0.1$^c$  &\hspace{-0.4cm} 0.1$^c$  &\hspace{-0.4cm} 0.1$^c$ &  -			 \\
\hspace{-0.15cm}$ Fe^d_{norm}$   &\hspace{-0.4cm} 1.2(7)    &\hspace{-0.4cm}  1.1(3)   &\hspace{-0.4cm} 0.8(1)   &\hspace{-0.4cm} 1.9(2)  &  -			 \\
\hspace{-0.2cm} EW (eV)          &\hspace{-0.4cm}  99(10)   &\hspace{-0.4cm}  101(30)  &\hspace{-0.4cm} 95(50)   &\hspace{-0.4cm} 160(80) &  -			 \\

\hspace{-0.3cm} abs.Flux$^{e}$   &\hspace{-0.4cm} 1.06(2)   &\hspace{-0.4cm} 1.08(2)   &\hspace{-0.4cm} 0.98(3)  &\hspace{-0.4cm} 0.62(3) &\hspace{-0.4cm} 0.0020(5)   \\
\hspace{-0.3cm} unab.Flux$^{e}$  &\hspace{-0.4cm} 1.24(3)   &\hspace{-0.4cm} 1.25(3)   &\hspace{-0.4cm} 1.03(5)  &\hspace{-0.4cm} 0.67(4) &\hspace{-0.4cm} 0.0020(5)   \\
C-stat                 		 &\hspace{-0.4cm} 866.4     &\hspace{-0.4cm} 878.7     &\hspace{-0.4cm} 632.5	 &\hspace{-0.4cm}  495.2  &\hspace{-0.4cm} 184.7       \\
     (d.o.f.)                    &\hspace{-0.4cm} (847)     &\hspace{-0.4cm} (893)     &\hspace{-0.4cm} (659)	 &\hspace{-0.4cm} (606)   &\hspace{-0.4cm} (162)      \\
\hline

\end{tabular}

$^{a}$ $\times10^{22}$\,cm$^{-2}$.  \\
$^{b}$ $\times10^{-2}$ photons keV$^{-1}$ cm$^{-2}$ s$^{-1}$ at 1\,keV.  \\
$^c$ Fixed.\\
$^d$   $\times10^{-3}$ photons cm$^{-2}$ s$^{-1}$ in the line. \\
$^{e}$ $\times10^{-9}$ erg cm$^{-2}$\,s$^{-1}$ in the 0.3--10\,keV  energy  band.\\
$^*$ See the text for additional results fixing the $N_{\rm H}$ parameter. \\
 \label{spec_tab}  
\end{table}
%%%%%%%%%%

\subsubsection{Timing Analysis and Results}\label{timing}

Figure\,\ref{plot_log} shows the most recent  X-ray activity of \sax\, observed  by \swift/BAT. The light-curve (binned using the  12.67\,d orbital period) shows the common feature for this source, a pre-outburst peak or flare anticipating all type II outbursts and lasting a few orbits. Since 2007 the peak intensity of these flares has decreased over time, while main outbursts reached approximately the same magnitude and duration ($\sim$7 months).  As seen in the optical band, the time span between X-ray type II outbursts was notably different, i.e.\  $\sim$2.7 yr between 2007 and 2010, and $\sim$0.7 yr between 2010 and 2012.  Periodic X-ray activity was not observed during quiescence, in agreement with previous results during quiescent states \citep[e.g.][]{baykal02, reig05, camero07, kiziloglu09, reig10}.  

To study the evolution of the pulse frequency of \sax~we started our search for pulsations in the data set from 2007.  For the frequency extraction we used the Fourier analysis module included in the \texttt{Period04} software package.\footnote{http://www.astro.univie.ac.at/dsn/dsn/Period04} \citep{Lenz05}  The Fourier analysis in \texttt{Period04}  is based on a discrete Fourier transform algorithm. The uncertainties in the frequency determinations are computed from Monte Carlo simulations.

The main problem we faced was the presence of red noise contaminating the space adjacent to the source's pulse period, making its precisely determination very difficult.  To address this issue we limited our search to a narrow frequency range centred on the values reported by \cite{ducci08} ($\sim$2.835 mHz) using \textit{INTEGRAL} data from the same period. We then fitted a sinusoid to a combination of the observed frequency and its second harmonic to obtain the best estimate for each XRT observation. This process yielded consistent frequency determinations of 2.834$\pm$0.022\,mHz, 2.835$\pm$0.011\,mHz, 2.839$\pm$0.002\,mHz, and 2.832$\pm$0.004\,mHz for the four observations from 2007 (see also Table~\ref{log_xrt}), respectively. The corresponding pulse periods are 352.8$\pm$2.7\,s, 352.7$\pm$1.3\,s, 352.2$\pm$0.2\,s and 353.0$\pm$0.5\,s.

\begin{figure}
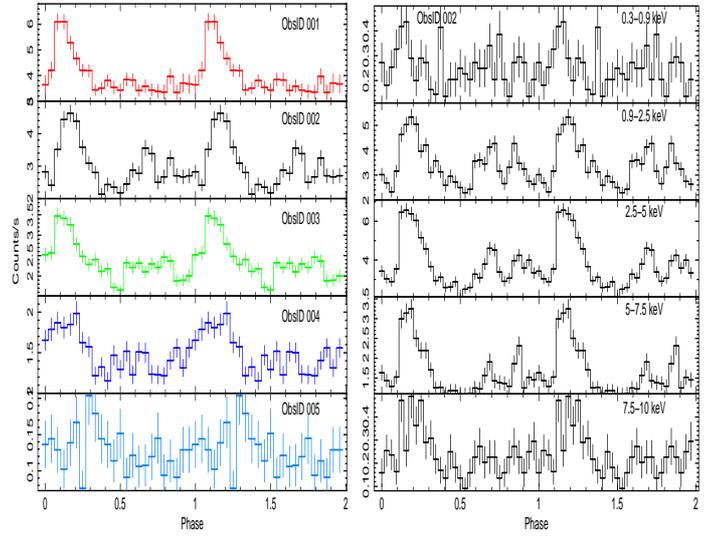

\hspace{-0.2cm}\hbox{
\rotatebox{-90}{\includegraphics[width=7cm,height=4.5cm]{figurota6_1.eps}}
\rotatebox{-90}{\includegraphics[width=7cm,height=4.5cm]{figurota6_2.eps}}
}
\caption{\textbf{Left}. \swift-XRT pulse profiles in the 0.3--10\,keV energy band for all the observations. \textbf{Right.} Pulse profiles (in counts s$^{-1}$) in different energy bands for the observation at higher luminosity - ObsID 002 (energy increasing from top to bottom).}
\label{profiles}
\end{figure}

In addition, we proceeded with the folding of the source events using our best frequency estimate obtained for each observation. The left panel of Fig.~\ref{profiles} shows  the pulse profiles during the 2007 flare (epoch increasing from top to bottom). The profiles showed a strong peak at phases $\sim$0-0.4 (becoming  wider in observations at lower luminosities) plus secondary weaker structures (here only found in observations ObsID 002 and 003) covering the rest of the phases and decreasing as the energy increases (see Fig.~\ref{profiles}).  In ObsID 002, although the two most prominent secondary peaks at phases $\sim$0.6--0.8 and $\sim$0.8--1 decreased as the energy increases, a small portion at phases 0.8--0.9 remained noticeable. On the other hand, all these weaker structures varied with time and seemed not to be correlated with luminosity. This confirms the known highly variable pulse profile behaviour of \sax~\citep{camero07}, whereas a good description of the processes responsible for the complex shape at low energies is still missing.  

\begin{figure*}
\includegraphics[width=18.5cm,height=8cm]{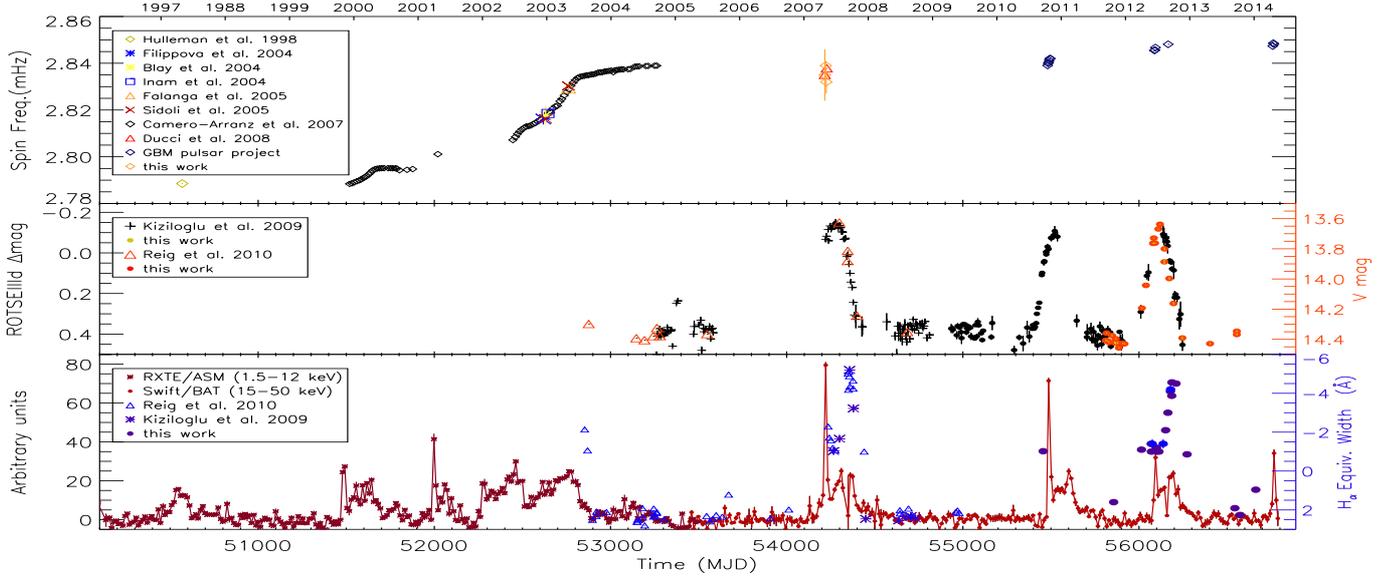}
\caption{\textbf{Top}. Long-term spin frequency history of \sax~ since its discovery in 1997. To visualize this plot more clearly we do not display the results from Baykal et al.\ (2002, 2007), our first XRT frequency determination from Table~\ref{log_xrt}, and some values from Sidoli et al.\ (2005).  \textbf{Middle}. Long-term optical lightcurve of this source. \textbf{Bottom}. History of the outbursts during the same period. The peak intensities of the different outbursts are only  intended to be illustrative of the times and types of events since they have not been corrected for the different energy bands. The long-term evolution of the \halfa EW (blue  triangles, stars, and filled circles) is also overplotted.}
\label{long_term}
\end{figure*}

\textit{Fermi}/GBM detected pulsations from SAX J2103.5+4545 at the beginning of the flare on 2010 October 12 (MJD 55481.4), at a frequency of 2.8390(3) mHz (P$_{spin}$=352.23(4)\,s).  The spin frequency increased at a rate of 2.3146912(1)$\times$10$^{-12}$ Hz s$^{-1}$ ($\dot{P}\sim-2.9\times10^{-7}$ s\,s$^{-1}$) during the flare from October 12 to 27 (see Fig.~\ref{long_term}). Between the 2010 and 2012 X-ray outbursts the spin frequency appears to increase smoothly at a rate of 0.66367141(3)$\times$10$^{-13}$ Hz s$^{-1}$ ($\dot{P}\sim-8.2\times10^{-9}$ s\,s$^{-1}$). On the contrary, between  2012 and 2014 this source spun down at a rate of $\sim1.5\times$10$^{-13}$ Hz s$^{-1}$ ($\dot{P}\sim1.8\times10^{-9}$ s\,s$^{-1}$). Then SAX J2103.5+4545 spun up  at a rate of 
$\sim$4.9$\times$10$^{-12}$ Hz s$^{-1}$ ($\dot{P}\sim$-6.1$\times10^{-7}$ s\,s$^{-1}$) during the first 3 days of the flare occurred in 2014.

In the present study, no pulsations were detected in the XRT observation during quiescence from 2012 May 2. One month later, \textit{Fermi}/GBM detected \sax~ at a frequency of 2.8454(3) mHz (P$_{spin}$=351.44(4)\,s) on 2012 June 9 (MJD 56087.6). From 2012 June 9 to August 26 the spin-up rate was  3.96(14) $\times$10$^{-13}$ Hz s$^{-1}$. We made an attempt to fold the source events using the GBM pulse detection and found that a single peak component might be the main feature (see bottom panel of Fig.~\ref{profiles}).  %However, the very low statistics prevent us from reaching any firm conclusion. 

The modulation amplitude of the 0.3--10\,keV pulses presented in the left panel of Fig.~\ref{profiles} can be measured using a pulse fraction defined as PF=F$_{max}$--F$_{min}$/F$_{max}$+F$_{min}$, where F$_{max}$ and F$_{min}$ are the maximum and minimum 
flux in the pulse profile, respectively. We found PF values of 33(4)$\%$, 41(6)$\%$, 37(5)$\%$, 31(6)$\%$, and 64(35)$\%$\,
($<$2$\sigma$) for ObsID 001 to 005,  respectively (see also Tab.~\ref{log_xrt}). Since this definition may add any noise present in the 
light curve to the true pulse fraction, thereby tending to increase the resulting measurement,  we used a more robust estimate of the PF, 
that is, the rms pulsed fraction defined as  PF$_{rms}$=$1/\bar{y}\,[1/n \sum\limits_{i=1}^n (y_{i}-\bar{y})^{2} -\sigma^2]^{1/2}$, where 
$n$ is the number of phase bins per cycle, $y_{i}$ is the number of counts in the $i$th phase bin, $\sigma_i$ is the error on $y_{i}$ and 
$\bar{y}$  is the mean number of counts in the cycle. We obtained PF$_{rms}$  values of 19(5)$\%$, 23(4)$\%$, 20(3)$\%$, 14(2)$\%$ for 
ObsID 001 to 004, respectively. However, for the observation during quiescence this definition did not yield a real solution. Finally, 
regarding the evolution of the PF$_{rms}$ with energy we obtained values of 21(3)$\%$, 28(5)$\%$, 41(6)$\%$, 57(4)$\%$, and 39(2)$\%$ for ObsID 002 in the 0.3--0.9\,keV, 0.9--2.5\,keV, 2.5--5\,keV, 5--7.5\,keV, and 7.5--10\,keV  energy bands, respectively. 

Furthermore, we studied the presence of the transient 22.7\,s QPO detected in an \textit{XMM-Newton} observation from 2003 January 6 \citep{inam04}. We did not find any QPO around 0.044 Hz in the power spectral density of the 0.3--10\,keV source events for any of the XRT observations.  A single power-law provided a good fit to the power spectra from 2007. The power-law indexes were 1.8(1), 1.4(2), 1.3(6) and 1.9(6) for ObsID 001 to 004, respectively. These values are all practically compatible within the uncertainties, and also with the values obtained by Reig et al.\ (2010) with \rxte/PCA during SAX\,J2103.5+4545's high state.

%%%%%  PULSE PROFILES 
%%%%%%%%%%%%%%%%%%%%%%%%

%%%%    long-term
%%%%%%%%%%%%%%%%%%%%%%%%

\section{Discussion}  \label{discuss}

\subsection{Long-term Be-disk/Neutron Star Interaction}
 
In this study we have confirmed the correlation between BV variability and X-ray intensity for \sax. Furthermore, we have found that this correlation extends to longer wavelengths in the IR domain. The optical/IR brightening of the Be star, besides moderate  values of the H$_\alpha$ line EW in emission, seems to precede the appearance of episodes of high X-ray luminosity (and neutron star spin-up). This behaviour has been observed in other BeXRBs; for instance, in EXO\,2030+375 \citep{wilson08}. 

Figures \ref{opt_ir} and \ref{long_term} 	corroborate that the optical outbursts of SAX\,J2103.5+4545 started about 3 months before the X-ray activity \citep[see also][]{kiziloglu09}. These episodes may be explained as mass ejection events from the donor star  that will provide the material for the Be disc formation \citep[see e.g.][for A\,0535+26]{yan12}. In this study we found the same type of behaviour in the IR for observations from 2012, interpreted as reprocessing in the Be disc.  The brightening events of 2010 and 2012 lasted for about 8/9 months, and so too probably did the one in 2007. The current activity exhibited by SAX\,J2103.5+4545 resembles the behaviour displayed in 2012. The model of \citet{okazakinegueruela01} ascribes the origin of X-ray outbursts to the truncation of the Be circumstellar disc at a resonance radius between the disc Keplerian velocity and the orbit of the neutron star.  Whether or not type I or II outbursts are produced depends on the eccentricity of the system as the disc expands and contracts between different resonant radii. This would explain the correlation between the disc status as inferred from H$\alpha$ EW and profile, and the presence or otherwise of X-ray outbursts in SAX\,J2103.5+4545. 

The relatively small peak value of the H$\alpha$ line EW in emission indicates a small Be disc truncated by the neutron star through its short orbit. The distance between the V and R peaks of the H$_\alpha$ line can be regarded with the measure of the H$_\alpha$ emitting 
region of the disc for a Be star \citep{huang72}: $(\Delta\,V/2\,v\,sin\,i)= (R_{disk}/R_{*})^{-j}$, where j=0.5 for Keplerian rotation 
(j=1 from the angular momentum conservation), R$_{disk}$ is the radius of H$_\alpha$  emitting region and R$_{*}$ denotes the radius 
of the central star. Taking the average value of $\Delta$V (381.12 km\,s$^{-1}$) and the previously found rotational velocity for this 
source (see Sect.~\ref{halpha}), the size of the H$_\alpha$ emitting region of SAX\,J2103.5+4545 is found to be R$_{disc}\sim$2.35 R$_{*}$, compatible with Reig et al.\ (2010). Furthermore,  the radius of the Be disc of SAX\,J2103.5+4545 was found to be similar to the size of the Roche lobe and similar to the periastron distance (Reig et al.\  2010), which probably propitiated  the formation of sudden and strong X-ray flares when the neutron star made direct contact with the Be disc during periastron passages. 

\cite{Kriss83} proposed an alternative view of the enhancement of the optical activity for the Be/X-ray binary system 4U\,0115+63 during an outburst in 1980. They suggested that the increase of optical light was created by material already transferred from the Be star to the vicinity of the neutron star,  and that it was radiating due to viscous heating in an accretion disc.  In the case of SAX\,J2103.5+4545, no X-rays were observed prior to 2012 May/June, so the transferred material must have been stored in an accretion disk since 2012 March. The delay of $\sim$90--100 days before the X-ray outburst would then represent the time scale for the material to move inward through the accretion disk to the neutron star surface under the action of viscous forces.  We can estimate the viscosity parameter $\alpha$ in the steady-state $\alpha$-disc model \citep{Shakura73, Novikov73}. The timescale given by \cite{Novikov73} is\linebreak
 $\delta\,t = 210\alpha^{-4/5} (M_*/1\,M_{\odot})^{1/4}(\dot{M}/10^{-10}M_{\odot}$ yr$^{-1}$)$^{-3/10}$ (R$_d$/10$^{11}$\,cm)$^{5/4}$ days, 
where M$_*$ is the mass of the neutron star, $\dot{M}$ is the accretion rate, and R$_d$ is the outer radius of the accretion disk. We can estimate the accretion rate from the observed X-ray luminosity during the outburst. Assuming a distance range of 3.2--6.5 kpc, a 10\,km neutron star radius, a mass of 1.4 M$_{\odot}$ and an energy conversion  efficiency of 0.5 (Baykal et al.\ 2002), we obtain $\dot{M}$= 0.4--1.5$\times$10$^{-9}$M$_{\odot}$\,yr$^{-1}$.  Expressing the disc size in terms of the viscosity parameter (where $\alpha<$1) gives an upper limit of the disc size of the order of  R$_d$ = ($\delta\,t/A )^{4/5}$\,10$^{11}$ cm (with A=156 or 102 for each distance estimate, respectively), that is,  R$_d$ = 6--9$\times10^{10}$ cm. Although the delay between the optical/IR and X-ray outbursts can be explained in this context, it is still difficult to explain why the optical/IR outbursts lasted for $\sim$9 months, while the X-ray outburst only $\sim$6 months. If matter was present in the accretion disc for this entire time, the X-ray outburst should have had a similar duration to the optical/IR events. A possibility is that the accretion disc was not a stable feature in the system.  Instead, there might have been a sudden inflow of disc material onto the neutron star after the disc has reached some critical configuration, as suggested by \cite{Kriss83} for 4U\,0115+63. This has been predicted in accretion discs in white dwarf systems, although not in neutron star systems.  For SAX\,J2103.5+4545 maybe only part of the material was initially transferred to the vicinity of the neutron star.  This scenario may be supported by the evolution of the H$_{\alpha}$ EW. From the fourth panel of Fig.~\ref{opt_ir} we see that a very small Be disc was still present and showed a stable size (for about 4 months) at the same time that the maxima of the optical/IR brightening episodes were reached.  After this period, however,  instead of fading with the optical/IR magnitudes,  the Be disc suddenly started growing  doubling its size up to its maximum of $\sim$1.7$\times$10$^{12}$ cm in $\sim$3 months. This is difficult to reconcile in terms of this alternative scenario. Therefore, we conclude that most likely  after a mass ejection event a Be disc was gradually formed, and then part of the matter was eventually transferred to the neutron star  giving rise to the X-ray outburst.

The present study restricts the time scale upper limit of the creation and disintegration of the Be disc to  $\sim$10 months,  the shortest known for BeXB due to its narrow orbit. From Table~\ref{ew_tab} we see that in 2012 the H$\alpha$ line EW in emission reached its maximum in $\sim$6.5 months (from MJD 56013.89--56210.02). Since the X-ray activity was triggered around MJD 56050 ($\sim$36 days after the H$\alpha$ was found in emission), this means that it most probably took no more than those $\sim$6.5 months for the disc to grow.  Two months later, in 2012 December (MJD 56272.2) the disc was still present but shortened. Although we found this line in absorption in the next available data in 2013 September (MJD 56543.79), from the X-ray observations we can infer that the transient disc was already not present $\sim$1.6 months later (around MJD 56320).  Analysis of  data from 2007 led Reig et al.\ (2010) to reach similar conclusions and find a time scale of $\sim$1.3--1.5 yr. 

This source is also particular because of the variability observed in the strength and shape of the H$\alpha$ line (see Fig.~\ref{ew_evol}).  Asymmetric line profiles may arise owing to one-armed density waves in the circumstellar disc, as suggested by the global disc oscillation model \citep{okazaki97, papaloizou06}.  Furthermore, cyclic changes in the V/R ratio  observed in many stars were also found to be consistent  with the presence of a one-armed density wave in the disc precessing around the central star \citep{porter03}. In this study, we observed rapid V/R ratio variations in SAX J2103.5+4545 (see Fig.~\ref{VR}); however, our sample did not allow us confidently  to predict a cyclic pattern. Furthermore,  rapid changes in line profile types are explained as the result of dynamical effects due to the misalignment of the disc orbital plane and the stellar equator \citep[see][]{porter03}. Recent studies propose that density changes in the disc thermal structure may be also the origin \citep[and references therein]{silaj10}. 

%\vspace*{-0.1cm}
\subsection{X-ray Behaviour}

 The spectral variability of \sax~ has been well studied by different authors, confirming the softening of this source as the flux decreased. Baykal et al.\ (2007) suggested that this type of spectral softening with decreasing flux  was mainly a consequence of mass accretion rate change. In particular, Reig et al.\ (2010, and references therein) explained that increased mass accretion rates are expected to result in harder X-ray spectra due to Comptonization processes. From the \textit{Swift}/XRT observations we saw a dramatic decrease of the X-ray luminosity of \sax\, from 2007 to 2012. Assuming a 6.5\,kpc distance (Reig et al.\ 2004), the luminosity dropped from $\sim$7 to  $\sim$0.01$\times10^{36}$\,erg s$^{-1}$.  Similar values have been previously observed at high states \citep{galis07, baykal07,camero07, kiziloglu09, reig10}.   The quiescent luminosity observed from this source with \textit{RXTE}/PCA is of the order of our XRT determination.

SAX\,J2103.5+4545 has shown a spin frequency increase since 1997 (see Fig.~\ref{long_term}) occasionally interrupted by short spin-down intervals (e.g.\ middle panel of fig.\ 12 of Camero-Arranz et al.\ 2007), corresponding to low accretion rates. It is to be noted that this source has  always been detected in all the available \textit{RXTE}/PCA observations from  2000 to 2004. The longest spin-up period  occurred between 2002 and 2004 (MJD 52445.90--52707.37), the most active so far, at a net rate of $\dot{\nu}\sim$1.3$\times$10$^{-12}$ Hz s$^{-1}$ ($\dot{P}\sim-8.2\times10^{-8}$ s\,s$^{-1}$). A torque/luminosity correlation was inferred from these observations, suggesting the presence of an accretion disc around the neutron star. This was later confirmed by the detection of a QPO in the \textit{XMM-Newton} data \citep{inam04}. All our pulse frequency measurements of SAX J2103.5+4545 during the 2007 X-ray flare are compatible within the uncertainties, and also with the values reported by \citep{ducci08} using \textit{INTEGRAL} data. Ducci et al.\ (2008) detected a similar spin-up trend for 12 days during the 2007 flare with $\dot{\nu}\sim$2.7$\times$10$^{-12}$ Hz s$^{-1}$ ($\dot{P}\sim-3.4\times10^{-7}$ s\,s$^{-1}$) . On the other hand,  our uncertainties do not allow us to distinguish any significant spin-up/down tendency. \textit{Fermi}/GBM observed a  spin-up rate  ($\sim$2.3$\times$10$^{-12}$ Hz s$^{-1}$; $\dot{P}\sim-2.9\times10^{-7}$ s\,s$^{-1}$) during approximately the 15 days that the 2010 flare lasted, and a similar spin-up ($\sim$2.6$\times$10$^{-12}$ Hz s$^{-1}$; $\dot{P}\sim-3.2\times10^{-7}$ s\,s$^{-1}$)  within the first 6 days of the 2012 flare. During the first 3 days of the flare occurred in 2014 the spin-up reached a value of $\sim$4.9$\times$10$^{-12}$ Hz s$^{-1}$ ($\dot{P}\sim-6.1\times10^{-7}$ s\,s$^{-1}$). These spin-up/X-ray flux correlations (see also Fig.~\ref{long_term}) suggest that an accretion disc around the neutron star was also present in the X-ray events from 2007 until 2014.

Photometric and spectroscopic observations performed in 2004 (Reig et al.\ 2005) during the faint state, showed that SAX\,J2103.5+4545 continued to emit X-rays even after having completely lost the Be disc. A gradual flux decline was also seen during quiescence after the type II outburst in 2004, yet the frequency continued to increase (see Fig~\ref{long_term}). This was unusual, but it has been seen in other wind-fed systems \citep{bildsten97}.  It was then proposed that, since SAX\,J2103.5+4545 is located in the region of wind-fed supergiant binaries in the P$_{spin}$--P$_{orb}$ or Corbet diagram, accretion from the stellar wind of the companion could be the origin of the observed luminosity. On the other hand, the random X-ray variability typical of wind-fed systems due to wind inhomogeneities is not present in \sax.~ Reig et al.\ (2010) proposed as the dominant accretion wind the more stable equatorial low-velocity high-density wind characteristic of BeXB. 

The evolutionary sequence that a neutron star will follow as it spins down from periods which are initially very short (P$<<$1 s) to moderate (P$\gtrsim$10 s) or long  (P$\gtrsim$100 s) periods,  have been discussed in several works \citep[see e.g.][]{davies_pringle81, henrichs83}.  Four distinct phases of neutron star spin-down in close binary systems have been proposed to be applicable to a particular range of rotation periods: pulsar-like, very rapid rotator, supersonic propeller and subsonic propeller (Davies \& Pringle, 1981). Therefore,  long period neutron stars, as SAX J2103.5+4545,  achieve the most effective braking during the subsonic propeller regimen, in which  the rotational rate decelerates due to the interaction between the magnetosphere and the surrounding hot, quasi-static plasma envelope \citep{Ikhsanov06}. While  SAX\,J2103.5+4545 was in a low state, between the 2004 December and 2007 April outbursts, Ducci et al.\ (2008) measured a spin-down of $\dot{\nu}\sim$\linebreak -0.4$\times$10$^{-13}$ Hz s$^{-1}$ ($\dot{P}\sim5.5\times10^{-9}$ s\,s$^{-1}$).  \textit{RXTE}/PCA observations from very low luminosity periods show that this source span down at similar rates (from $\sim$ -2 to -0.2$\times$10$^{-13}$ Hz s$^{-1}$) \citep{baykal07, camero07}.  Based on our observations during  the same quiescent period,  $\dot\nu$\, $\sim$\,-0.5$\times$10$^{-13}$ Hz s$^{-1}$ ($\dot{P}\sim5.9\times10^{-9}$ s\,s$^{-1}$).  Assuming that in quiescence SAX\,J2103.5+4545 enters the subsonic propeller regime, it is expected to spin down at a rate of  $\dot\nu_{sub}=-4\pi\nu^2\mu^2\,(GM)^{-1}I^{-1}$, where  $\mu$ is the neutron star magnetic moment, $M$ the mass (1.4 M$\odot$) and $I$ the moment of inertia \citep{henrichs83}. If we select the two reported values of the magnetic field for SAX\,J2103.5+4545 - that is, B=1.65$\times10^{12}$\,G (Sidoli et al.\ 2005) and 16.5 $\times10^{12}$\,G (Baykal et al.\ 2007) -  then $\dot\nu_{sub}\sim$ -0.015 and -1.5 $\times$10$^{-13}$ Hz s$^{-1}$, respectively. Therefore, comparing these results with  our observations, the behaviour of this source during quiescence may be understood only by assuming B=16.5$\times10^{12}$\,G.

\section{Summary}

In this article we have presented the results of our multiwavelength campaign for the Be/X-ray binary system SAX\,J2103.5+4545.  We have performed spectral and photometric temporal analysis in order to investigate the transient behaviour exhibited by this source since 2007. These new observations were put into the context of historical data and discussed in terms of the neutron star Be-disc interaction.  X-ray spectral analysis of observations during active periods and quiescence support previous findings for this system. In particular, the photon indexes were consistent with SAX\,J2103.5+4545 being at high/low luminosity states. Pulsations were found only in the data from 2007 but not during quiescence. Our timing analyses results also support the presence of a transient accretion disc around the neutron star during major periods of activity (each event lasting $\sim$6/7 months since 2007).  The optical outbursts in 2010 and 2012 lasted for about 8/9 months (as did probably the one in 2007 and will probably do the one in 2014), and were most probably due to mass ejection events from the Be star. Thanks to our long-term IR monitoring of this source, a correlation between the IR variability and the X-ray intensity was found. The IR enhancement episode of 2012 extended for the same period of time as in the optical band. The optical/IR outbursts started about 3 months before the triggering of the X-ray activity. However, the optical/IR brightness and the H$\alpha$ EW were anti-correlated; that is, the maximum of the EW of this line was reached during the decline of the brightness of the BVJHKs magnitudes.  We confined the disc formation/disintegration process to within about 10 months. We only observed H$\alpha$ line profiles in absorption at the beginning and end of the optical/IR activity, an indication that no disc was present around the Be star. We observed fast  H$\alpha$ line variability, and asymmetric double and single profiles in emission during the evolution of the Be disc. This behaviour might be explained in terms of one-armed density waves in the circumstellar disc.

\textbf{Acknowledgments}. This article is partially based on service observations made with the IAC80 and TCS telescopes operated on the island of Tenerife by the Instituto de Astrof\'{\i}sica de Canarias (IAC) in the Spanish Observatorio del Teide. The present work is also based on observations made with the Nordic Optical Telescope, operated by the Nordic Optical Telescope Scientific Association at the Observatorio del Roque de los Muchachos (IAC), La Palma, Spain. The Albireo spectrograph at the 1.5-m telescope is operated by the Instituto de  Astrof\'{\i}sica de Andaluc\'{\i}a at the Sierra Nevada Observatory. M.\"{O}. A. acknowledges support from T\"{U}B\.{I}TAK, The Scientific and Technological Research Council of Turkey, through the research project 106T040. We thank T\"{U}B\.{I}TAK and ROTSE collaboration for partial support in using the RTT 150 and ROTSEIIId Telescopes with project numbers TUG-RTT150.08.45, 12ARTT150-264-1 and ROTSE-40. We also thank  \"{U}. K{\i}z{\i}lo{\v g}lu for providing all the available ROTSE data on SAX J2103.5+4545 for this study.  We thank H. Bilal\"{O}zcan for the most recent optical spectroscopic observations obtained with the RTT150.  We appreciate to A. Papitto for useful discussions about spindown processes in neutron stars systems. We also acknowledge Terry Mahoney for revising the English text. This study made use of data supplied by the UK \textit{Swift} Science Data Centre at the University of Leicester. The work of J.G.S. is supported by the Spanish Programa Nacional de Astronom\'{\i}a y Astrof\'{\i}sica under contract AYA2012-39246-C02-01. E. N. acknowledges a VALi+d postdoctoral grant from the Generalitat Valenciana and was supported by the Spanish Ministry of Economy and Competitiveness under contract AYA 2010-18352. MCG is supported by the European social fund within the framework of
"Support of inter-sectoral mobility and quality enhancement of research teams at Czech Technical University in Prague“,
CZ.1.07/2.3.00/30.0034. A. C. was supported by the AYA2012-39303, SGR2009-811 and iLINK2011-0303 grants.

%\bibliographystyle{bibstyles/astron}
%\bibliography{saxj2103}

\begin{thebibliography}{}


\bibitem[\protect\astroncite{{Akerlof} et~al.}{2003}]{akerlof03}
{Akerlof}, C.~W., {Kehoe}, R.~L., {McKay}, T.~A., et al.: 2003,
\newblock {\em \pasp} {   115}, 132

\bibitem[\protect\astroncite{{Anders} and {Grevesse}}{1989}]{anders_grevesse89}
{Anders}, E. and {Grevesse}, N.: 1989,
\newblock {\em \gca} {   53}, 197

\bibitem[\protect\astroncite{{Arnaud}}{1996}]{arnaud96}
{Arnaud}, K.~A.: 1996,
\newblock in G.~H. {Jacoby} and J. {Barnes} (eds.), {\em Astronomical Data
  Analysis Software and Systems V}, Vol. 101 of {\em Astronomical Society of
  the Pacific Conference Series}, p.~17

\bibitem[\protect\astroncite{{Balucinska-Church} and
  {McCammon}}{1992}]{balucinska92}
{Balucinska-Church}, M. and {McCammon}, D.: 1992,
\newblock {\em \apj} {   400}, 699

\bibitem[\protect\astroncite{{Baykal} et~al.}{2007}]{baykal07}
{Baykal}, A., {Inam}, S.~{\c C}., {Stark}, M.~J., et al.: 2007,
\newblock {\em \mnras} {   374}, 1108

\bibitem[\protect\astroncite{{Baykal} et~al.}{2000}]{baykal00}
{Baykal}, A., {Stark}, M.~J., and {Swank}, J.: 2000,
\newblock {\em \apjl} {   544}, L129

\bibitem[\protect\astroncite{{Baykal} et~al.}{2002}]{baykal02}
{Baykal}, A., {Stark}, M.~J., and {Swank}, J.~H.: 2002,
\newblock {\em \apj} {   569}, 903

\bibitem[\protect\astroncite{{Bildsten} et~al.}{1997}]{bildsten97}
{Bildsten}, L., {Chakrabarty}, D., {Chiu}, J., et al.:1997,
\newblock {\em \apjs} {   113}, 367

\bibitem[\protect\astroncite{{Blay} et~al.}{2004}]{blay04}
{Blay}, P., {Reig}, P., {Mart{\'{\i}}nez N{\'u}nez}, S., et al.: 2004,
\newblock {\em \aap} {   427}, 293

\bibitem[\protect\astroncite{{Blay} et~al.}{2006}]{blay06}
{Blay}, P., {Negueruela}, I., {Reig}, P., et al.: 2006,
\newblock {\em \aap} {   446}, 1095

\bibitem[\protect\astroncite{{Burrows} et~al.}{2005}]{burrows05}
{Burrows}, D.~N.,{Hill}, J.~E.,{Nousek}, J.~A., et al.:2005,
\newblock {\em \ssr} {   120}, 165


\bibitem[\protect\astroncite{{Camero-Arranz} et~al.}{2010}]{camero10}
{Camero-Arranz}, A., {Finger}, M.~H., {Ikhsanov}, N.~R., et al.: 2010,
\newblock {\em \apj} {   708}, 1500


\bibitem[\protect\astroncite{{Camero-Arranz} et~al.}{2012}]{camero12}
{Camero-Arranz}, A., {Nespoli}, E., {Gutierrez-Soto}, J., and {Zurita}, C.:
  2012b,
\newblock {\em The Astronomer's Telegram} {   4187}, 1

\bibitem[\protect\astroncite{{Camero Arranz} et~al.}{2007}]{camero07}
{Camero Arranz}, A., {Wilson}, C.~A., {Finger}, M.~H., and {Reglero}, V.: 2007,
\newblock {\em \aap} {   473}, 551

\bibitem[\protect\astroncite{{Davies} and {Pringle}}{1981}]{davies_pringle81}
{Davies}, R.~E. and {Pringle}, J.~E.: 1981,
\newblock {\em \mnras} {   196}, 209


\bibitem[\protect\astroncite{{Ducci} et~al.}{2008}]{ducci08}
{Ducci}, L., {Sidoli}, L., {Paizis}, A., et al.: 2008,
\newblock in {\em Proceedings of the 7th INTEGRAL Workshop}

\bibitem[\protect\astroncite{{Evans} et~al.}{2009}]{evans09}
{Evans}, P.~A., {Beardmore}, A.~P., {Page}, K.~L., et al.: 2009,
\newblock {\em \mnras} {   397}, 1177

\bibitem[\protect\astroncite{{Falanga} et~al.}{2005}]{falanga05}
{Falanga}, M., {di Salvo}, T., {Burderi}, L., et al.: 2005,
\newblock {\em \aap} {   436}, 313

\bibitem[\protect\astroncite{{Filippova} et~al.}{2004}]{filippova04}
{Filippova}, E.~V., {Lutovinov}, A.~A., {Shtykovsky}, P.~E., et al.: 2004,
\newblock {\em Astronomy Letters} {   30}, 824

\bibitem[\protect\astroncite{{Finger} et~al.}{2009}]{finger09}
{Finger}, M.~H., {Beklen}, E., {Narayana Bhat}, P.,  et al.:
  2009,
\newblock {\em ArXiv e-prints}

\bibitem[\protect\astroncite{{Frank} et~al.}{2002}]{frank02}
{Frank}, J., {King}, A., and {Raine}, D.~J.: 2002,
\newblock {\em {Accretion Power in Astrophysics: Third Edition}}

\bibitem[\protect\astroncite{{Galis} et~al.}{2007}]{galis07}
{Galis}, R., {Beckmann}, V., {Bianchin}, V., et al.: 2007,
\newblock {\em The Astronomer's Telegram} {   1063}, 1

\bibitem[\protect\astroncite{{Gehrels} et~al.}{2004}]{gehrels04}
{Gehrels}, N.,{Chincarini}, G.,{Giommi}, P., et al.:2004,
\newblock {\em \apj} {   611}, 1005

\bibitem[\protect\astroncite{{Hanuschik}}{1989}]{hanuschik89}
{Hanuschik}, R.~W.: 1989,
\newblock {\em \apss} {   161}, 61

\bibitem[\protect\astroncite{{Hanuschik} et~al.}{1988}]{hanuschik88}
{Hanuschik}, R.~W., {Kozok}, J.~R., and {Kaiser}, D.: 1988,
\newblock {\em \aap} {   189}, 147

\bibitem[\protect\astroncite{{Henrichs}}{1983}]{henrichs83}
{Henrichs}, H. (ed.): 1983,
\newblock {\em {Accretion-Driven Stellar X-ray sources. Cambrige Univ. Press,
  393.}}

\bibitem[\protect\astroncite{{Huang}}{1972}]{huang72}
{Huang}, S.-S.: 1972,
\newblock {\em \apj} {   171}, 549

\bibitem[\protect\astroncite{{Hulleman} et~al.}{1998}]{Hulleman98}
{Hulleman}, F., {in 't Zand}, J.~J.~M., and {Heise}, J.: 1998,
\newblock {\em \aap} {   337}, L25

\bibitem[\protect\astroncite{{Ikhsanov} and {Choi}}{2006}]{Ikhsanov06}
{Ikhsanov}, N.~R. and {Choi}, C.-S.: 2006,
\newblock {\em Advances in Space Research} {  38}, 2901

\bibitem[\protect\astroncite{{{\.I}nam} et~al.}{2004}]{inam04}
{{\.I}nam}, S.~{\c C}., {Baykal}, A., {Swank}, J., and {Stark}, M.~J.: 2004,
\newblock {\em \apj} {   616}, 463


\bibitem[\protect\astroncite{{Kiziloglu} et~al.}{2010}]{kiziloglu10}
{Kiziloglu}, U., {Kiziloglu}, N., {Baykal}, A., and {Inam}, S.~C.: 2010,
\newblock {\em The Astronomer's Telegram} {   2925}, 1

\bibitem[\protect\astroncite{{K{\i}z{\i}lo{\v g}lu} et~al.}{2005}]{kiziloglu05}
{K{\i}z{\i}lo{\v g}lu}, {\"U}., {K{\i}z{\i}lo{\v g}lu}, N., and {Baykal}, A.:
  2005,
\newblock {\em \aj} {   130}, 2766

\bibitem[\protect\astroncite{{K{\i}z{\i}lo{\v g}lu} et~al.}{2009}]{kiziloglu09}
{K{\i}z{\i}lo{\v g}lu}, {\"U}., {{\"O}zbilgen}, S., {K{\i}z{\i}lo{\v g}lu}, N.,
  and {Baykal}, A.: 2009,
\newblock {\em \aap} {   508}, 895

\bibitem[\protect\astroncite{Konstantinova and Mokrushina}{2014}]{konstantinova_mokrushina14}
{Konstantinova}, T.~S., {Mokrushina}, A.~A.: 2014,
\newblock {\em The Astronomer's Telegram} {   6019}, 1

\bibitem[\protect\astroncite{{Krimm} et~al.}{2007}]{krimm07}
{Krimm}, H.~A., {Barthelmy}, S.~D., {Barbier}, L., et al.: 2007,
\newblock {\em The Astronomer's Telegram} {   1064}, 1

\bibitem[\protect\astroncite{{Krimm} et~al.}{2010}]{Krimm10}
{Krimm}, H.~A., {Barthelmy}, S.~D., {Baumgartner}, W., et al.: 2010,
\newblock {\em The Astronomer's Telegram} {   2928}, 1

\bibitem[\protect\astroncite{{Krimm} et~al.}{2013}]{krimm13}
{Krimm}, H.~A., {Holland}, S.~T., {Corbet}, R.~H.~D., et al.: 2013,
\newblock {\em \apjs} {   209}, 14

\bibitem[\protect\astroncite{{Kriss} et~al.}{1983}]{Kriss83}
{Kriss}, G.~A. and {Cominsky}, L.~R. and {Remillard}, R.~A.,et al.: 1983
\newblock {\em \apj} {   266}, 806

\bibitem[\protect\astroncite{{Lenz} and {Breger}}{2005}]{Lenz05}
{Lenz}, P. and {Breger}, M.: 2005,
\newblock {\em Communications in Asteroseismology} {   146}, 53

\bibitem[\protect\astroncite{{Lutovinov} et~al.}{2003}]{lutovinov03}
{Lutovinov}, A.~A., {Molkov}, S.~V., and {Revnivtsev}, M.~G.: 2003,
\newblock {\em Astronomy Letters} {   29}, 713

\bibitem[\protect\astroncite{{Manousakis} et~al.}{2007}]{Manousakis07}
{Manousakis}, A., {Reig}, P., and {Kougentakis}, A.: 2007,
\newblock {\em The Astronomer's Telegram} {   1085}, 1

\bibitem[\protect\astroncite{{Novikov} and {Thorne}}{1973}]{Novikov73}
{Novikov}, I.~D. and {Thorne}, K.~S.:1973,
\newblock {\em Black Holes (Les Astres Occlus), ed. C. Dewitt and B. S. Dewitt (New York, Gordon \& Breach)}, 343

\bibitem[\protect\astroncite{{Okazaki}}{1997}]{okazaki97}
{Okazaki}, A.~T.: 1997,
\newblock {\em \aap} {   318}, 548

\bibitem[\protect\astroncite{{Okazaki} and {Negueruela}}{2001}]{okazakinegueruela01}
{Okazaki}, A.~T. and {Negueruela}, I.: 2001,
\newblock {\em \aap} {   377}, 161

\bibitem[\protect\astroncite{{Papaloizou} and {Savonije}}{2006}]{papaloizou06}
{Papaloizou}, J.~C.~B. and {Savonije}, G.~J.: 2006,
\newblock {\em \aap} {   456}, 1097

\bibitem[\protect\astroncite{{Porter} and {Rivinius}}{2003}]{porter03}
{Porter}, J.~M. and {Rivinius}, T.: 2003,
\newblock {\em \pasp} {   115}, 1153

\bibitem[\protect\astroncite{{Reig} et~al.}{2004}]{reig04}
{Reig}, P., {Negueruela}, I., {Fabregat}, J., et al.: 2004,
\newblock {\em \aap} {   421}, 673

\bibitem[\protect\astroncite{{Reig} et~al.}{2005}]{reig05}
{Reig}, P., {Negueruela}, I., {Papamastorakis}, G., et al.: 2005,
\newblock {\em \aap} {   440}, 637

\bibitem[\protect\astroncite{{Reig} et~al.}{2010}]{reig10}
{Reig}, P., {S{\l}owikowska}, A., {Zezas}, A., and {Blay}, P.: 2010,
\newblock {\em \mnras} {   401}, 55

\bibitem[\protect\astroncite{{Sguera} et~al.}{2012}]{sguera12}
{Sguera}, V., {Drave}, S., {Goossens}, M., et al.: 2012,
\newblock {\em The Astronomer's Telegram} {   4168}, 1

\bibitem[\protect\astroncite{{Shakura} and {Sunyaev}}{1973}]{Shakura73}
{Shakura}, N.~I. and {Sunyaev}, R.~A.:1973
\newblock {\em \aap} {   24}, 337

\bibitem[\protect\astroncite{{Sidoli} et~al.}{2005}]{sidoli05}
{Sidoli}, L., {Mereghetti}, S., {Larsson}, S., et al.: 2005,
\newblock {\em \aap} {   440}, 1033

\bibitem[\protect\astroncite{{Silaj} et~al.}{2010}]{silaj10}
{Silaj}, J., {Jones}, C.~E., {Tycner}, C., et al.: 2010,
\newblock {\em \apjs} {   187}, 228

\bibitem[\protect\astroncite{{Staubert} et~al.}{2011}]{staubert11}
{Staubert}, R., {Pottschmidt}, K., {Doroshenko}, V., et al.: 2011,
\newblock {\em \aap} {   527}, A7


\bibitem[\protect\astroncite{{Steele} et~al.}{1999}]{steele99}
{Steele}, I.~A., {Negueruela}, I., and {Clark}, J.~S.: 1999,
\newblock {\em \aaps} {   137}, 147			   

\bibitem[\protect\astroncite{{Wilson} et~al.}{2008}]{wilson08}
{Wilson}, C.~A., {Finger}, M.~H., and {Camero-Arranz}, A.: 2008,
\newblock {\em \apj} {   678}, 1263

\bibitem[\protect\astroncite{{Yan} et~al.}{2012}]{yan12}
{Yan}, J., {Li}, H., and {Liu}, Q.: 2012,
\newblock {\em \apj} { 744}, 37

\pagebreak

\end{thebibliography}

\pagebreak

%%%%%%%%  OPtical photometry
\begin{table*}
\caption{\textbf{ON-LINE MATERIAL}. Log of the optical photometric observations of SAX J2103.5+4545 with the IAC80 telescope.}          
%\begin{center}                
\begin{tabular}{lccc} 
\hline\hline

 Date        &  MJD 	   &	  B	       &	 Vmag	   \\
   		  &    &   (mag)        &        (mag)	\\
   \hline\hline
  2011/09/08    &    55812.071  &  15.587$\pm$0.007  &  14.571$\pm$0.004   \\ 
  2011/09/12   &    55816.049  &  15.528$\pm$0.008  &  14.535$\pm$0.003   \\ 
  2011/09/18   &    55822.883  &  15.589$\pm$0.005  &  14.574$\pm$0.003   \\ 
  2011/09/24   &    55828.994  &  15.600$\pm$0.006  &  14.581$\pm$0.003   \\ 
  2011/10/13  &    55847.055  &  15.52$\pm$0.01  &  14.547$\pm$0.005   \\ 
  2011/10/27  &    55861.925  &  15.591$\pm$0.005  &  14.582$\pm$0.005   \\ 
  2011/10/27  &    55861.922  &   -		   &  14.580$\pm$0.003    \\ 
  2011/11/02   &    55867.990  &  15.57$\pm$0.01  &  14.585$\pm$0.007   \\ 
  2011/11/02   &    55867.992  &  15.60$\pm$0.01  &  14.563$\pm$0.003   \\ 
  2011/11/13  &    55878.875  &  15.599$\pm$0.009  &  14.583$\pm$0.004   \\ 
  2011/11/20  &    55885.819  &  15.609$\pm$0.005  &  14.608$\pm$0.003   \\ 
  2011/12/13  &    55908.871  &  15.603$\pm$0.006  &  14.585$\pm$0.003   \\ 
  2011/12/22  &    55917.841  &  15.574$\pm$0.005  &  14.588$\pm$0.003   \\ 
  2012/03/30   &    56016.221  &  15.471$\pm$0.006  &  14.412$\pm$0.003   \\ 
  2012/04/22   &    56039.218  &  15.384$\pm$0.004  &  14.300$\pm$0.002   \\ 
  2012/06/02    &    56080.187  &  15.195$\pm$0.004  &  14.093$\pm$0.002   \\ 
  2012/06/04    &    56082.195  &  15.195$\pm$0.006  &  14.092$\pm$0.003   \\ 
  2012/06/06    &    56084.166  &  15.150$\pm$0.004  &  14.067$\pm$0.002   \\ 
  2012/06/13   &    56091.144  &  15.196$\pm$0.006  &  14.092$\pm$0.003   \\ 
  2012/06/30   &    56108.939  &  15.148$\pm$0.008  &  14.022$\pm$0.003   \\ 
  2012/07/10   &    56118.926  &  15.159$\pm$0.006  &  13.999$\pm$0.003   \\ 
  2012/08/04    &    56143.970  &  15.28$\pm$0.02  &  14.184$\pm$0.015  \\ 
  2012/08/06    &    56145.018  &  15.25$\pm$0.01  &  14.120$\pm$0.006   \\ 
  2012/09/02    &    56172.113  &  15.35$\pm$0.01  &  14.266$\pm$0.006   \\ 
  2012/09/02    &    56172.110  &  15.37$\pm$0.01  &   -    \\ 
  2012/09/25   &    56195.944  &  15.47$\pm$0.01  &   -    \\ 
  2012/09/25   &    56195.945  &  15.43$\pm$0.02  &   -    \\ 
  2012/09/25   &    56195.947  &  15.45$\pm$0.02  &   -    \\ 
  2012/09/25   &    56195.948  &  15.44$\pm$0.01  &   -    \\ 
  2012/09/25   &    56195.952  &  15.45$\pm$0.01  &   -    \\ 
  2012/09/25   &    56195.953  &  15.44$\pm$0.02  &  14.389$\pm$0.006   \\ 
  2012/11/14  &    56245.918  &  15.56$\pm$0.01  &  14.559$\pm$0.007   \\ 
  2013/04/21   &    56403.281  &   -  &  14.587$\pm$0.006   \\ 
  2013/09/19   &    56554.946  &  15.48$\pm$0.01  &  14.526$\pm$0.005   \\ 
  2013/09/19   &    56554.947  &  15.53$\pm$0.01  &  14.542$\pm$0.005   \\ 

\hline
\end{tabular} \label{opt_phot}  
%\end{center}
\end{table*}

%%%%%%%%  IR photometry  
\begin{table*}
\caption{\textbf{ON-LINE MATERIAL}. IR photometric observations of SAX J2103.5+4545 with the TCS telescope.}                          
\begin{tabular}{llccc} 
\hline\hline
Date    &   MJD  & \hspace{-0.6cm} J &\hspace{-0.6cm}H & \hspace{0.1cm}Ks  \\
          &     & \hspace{-0.6cm}(mag)  &\hspace{-0.6cm}(mag) &(mag) \\
\hline\hline
  \hspace{-0.2cm}2011/09/11  &   \hspace{-0.2cm}55815.125  & \hspace{-0.3cm}11.99$\pm$0.07     & \hspace{-0.3cm}11.52$\pm$0.07   & \hspace{-0.3cm}11.36$\pm$0.07	 \\ 
\hspace{-0.2cm}2011/09/24  &   \hspace{-0.2cm}55828.999  & \hspace{-0.3cm} -  	       & \hspace{-0.3cm}11.51$\pm$0.07   & \hspace{-0.3cm}11.27$\pm$0.07   \\ 
\hspace{-0.2cm}2011/11/13  &   \hspace{-0.2cm}55878.906  & \hspace{-0.3cm}11.75$\pm$0.07     & \hspace{-0.3cm}11.50$\pm$0.07   & \hspace{-0.3cm}11.29$\pm$0.07	   \\ 
\hspace{-0.2cm}2011/12/03  &   \hspace{-0.2cm}55898.849  & \hspace{-0.3cm}11.77$\pm$0.07     & \hspace{-0.3cm}11.44$\pm$0.07   & \hspace{-0.3cm}11.26$\pm$0.07	    \\ 
\hspace{-0.2cm}2011/12/26  &   \hspace{-0.2cm}55921.838  & \hspace{-0.3cm}11.81$\pm$0.07     & \hspace{-0.3cm}11.47$\pm$0.07   & \hspace{-0.3cm}11.30$\pm$0.07	    \\ 
\hspace{-0.2cm}2012/04/10  &   \hspace{-0.2cm}56027.180  & \hspace{-0.3cm}11.33$\pm$0.07     & \hspace{-0.3cm}10.87$\pm$0.07   & \hspace{-0.3cm}10.52$\pm$0.07	    \\ 
\hspace{-0.2cm}2012/05/07  &   \hspace{-0.2cm}56054.201  & \hspace{-0.3cm}11.14$\pm$0.07     & \hspace{-0.3cm}10.49$\pm$0.07   & \hspace{-0.3cm}10.43$\pm$0.07	    \\ 
\hspace{-0.2cm}2012/05/29  &   \hspace{-0.2cm}56076.185  & \hspace{-0.3cm}10.98$\pm$0.07     & \hspace{-0.3cm}     - 	  & \hspace{-0.3cm}10.38$\pm$0.07   \\ 
\hspace{-0.2cm}2012/06/11  &   \hspace{-0.2cm}56089.201  & \hspace{-0.3cm}10.97$\pm$0.07     & \hspace{-0.3cm}10.59$\pm$0.07   & \hspace{-0.3cm}10.25$\pm$0.07	  \\ 
\hspace{-0.2cm}2012/06/15  &   \hspace{-0.2cm}56093.084  & \hspace{-0.3cm}10.99$\pm$0.07     & \hspace{-0.3cm}10.42$\pm$0.07   & \hspace{-0.3cm}     -	     \\ 
\hspace{-0.2cm}2012/06/17  &   \hspace{-0.2cm}56095.119  & \hspace{-0.3cm}10.96$\pm$0.07     & \hspace{-0.3cm}10.56$\pm$0.07   & \hspace{-0.3cm}10.27$\pm$0.07     \\ 
\hspace{-0.2cm}2012/07/21  &   \hspace{-0.2cm}56129.194  & \hspace{-0.3cm}10.85$\pm$0.07     & \hspace{-0.3cm}10.49$\pm$0.07   & \hspace{-0.3cm}10.19$\pm$0.07     \\ 
\hspace{-0.2cm}2012/07/29  &   \hspace{-0.2cm}56137.112  & \hspace{-0.3cm}10.96$\pm$0.03     & \hspace{-0.3cm}10.54$\pm$0.03   & \hspace{-0.3cm}10.112$\pm$0.008   \\ 
\hspace{-0.2cm}2012/07/31  &   \hspace{-0.2cm}56139.023  & \hspace{-0.3cm}10.953$\pm$0.005   & \hspace{-0.3cm}10.50$\pm$0.01   & \hspace{-0.3cm}10.23$\pm$0.03  \\ 
\hspace{-0.2cm}2012/08/02  &   \hspace{-0.2cm}56141.053  & \hspace{-0.3cm}10.927$\pm$0.009   & \hspace{-0.3cm}10.472$\pm$0.003 & \hspace{-0.3cm}-  \\ 
\hspace{-0.2cm}2012/08/02  &   \hspace{-0.2cm}56141.067  & \hspace{-0.3cm} -  	       & \hspace{-0.3cm}10.53$\pm$0.02   & \hspace{-0.3cm} 10.21$\pm$0.03	      \\ 
\hspace{-0.2cm}2012/08/06  &   \hspace{-0.2cm}56145.120  & \hspace{-0.3cm}10.934$\pm$0.011   & \hspace{-0.3cm}10.453$\pm$0.015 & \hspace{-0.3cm}10.15$\pm$0.02  \\ 
\hspace{-0.2cm}2012/08/13  &   \hspace{-0.2cm}56152.969  & \hspace{-0.3cm}10.964$\pm$0.004   & \hspace{-0.3cm}10.481$\pm$0.002 & \hspace{-0.3cm}10.16$\pm$0.04  \\ 
\hspace{-0.2cm}2012/09/08  &   \hspace{-0.2cm}56178.980  & \hspace{-0.3cm}11.160$\pm$0.008   & \hspace{-0.3cm}10.612$\pm$0.002 & \hspace{-0.3cm}10.25$\pm$0.03  \\ 
\hspace{-0.2cm}2012/10/01  &   \hspace{-0.2cm}56201.010  & \hspace{-0.3cm}11.354$\pm$0.008   & \hspace{-0.3cm}11.11$\pm$0.02   & \hspace{-0.3cm}10.782$\pm$0.008   \\ 
\hspace{-0.2cm}2012/10/14  &   \hspace{-0.2cm}56214.890  & \hspace{-0.3cm}11.83$\pm$0.03     & \hspace{-0.3cm}11.24$\pm$0.01   & \hspace{-0.3cm}10.84$\pm$0.04  \\ 
\hspace{-0.2cm}2012/12/16  &   \hspace{-0.2cm}56277.849  & \hspace{-0.3cm}11.780$\pm$0.003   & \hspace{-0.3cm}11.41$\pm$0.01   & \hspace{-0.3cm}11.22$\pm$0.02  \\ 
\hspace{-0.2cm}2013/04/22  &   \hspace{-0.2cm}56404.166  & \hspace{-0.3cm}11.824$\pm$0.002   & \hspace{-0.3cm}11.54$\pm$0.02   & \hspace{-0.3cm}11.35$\pm$0.02  \\ 
\hspace{-0.2cm}2013/06/30  &   \hspace{-0.2cm}56473.117  & \hspace{-0.3cm}11.81$\pm$0.06     & \hspace{-0.3cm}11.46$\pm$0.03   & \hspace{-0.3cm}11.38$\pm$0.07  \\ 
\hspace{-0.2cm}2013/09/11  &   \hspace{-0.2cm}56546.076  & \hspace{-0.3cm}11.85$\pm$0.03     & \hspace{-0.3cm}11.41$\pm$0.07   & \hspace{-0.3cm}11.29$\pm$0.08	  \\  
\hspace{-0.2cm}2013/11/10  &   \hspace{-0.2cm}56606.907  & \hspace{-0.3cm}11.81$\pm$0.03     & \hspace{-0.3cm}    -  	  & \hspace{-0.3cm} -  	 \\ 
\hspace{-0.2cm}2013/11/10  &   \hspace{-0.2cm}56606.913  & \hspace{-0.3cm} -  	       & \hspace{-0.3cm}11.48$\pm$0.01   & \hspace{-0.3cm}11.39$\pm$0.04	  \\
\hspace{-0.2cm}2014/11/10  &   \hspace{-0.2cm}56606.913  & \hspace{-0.3cm} -  	       & \hspace{-0.3cm}11.48$\pm$0.01   & \hspace{-0.3cm}11.39$\pm$0.04	  \\
\hspace{-0.2cm}2014/04/05  &   \hspace{-0.2cm}56752.182  & \hspace{-0.3cm} 11.31	$\pm$0.07 	   & \hspace{-0.3cm}10.87$\pm$0.07   & \hspace{-0.3cm}10.70$\pm$0.07	  \\
\hspace{-0.2cm}2014/04/05  &   \hspace{-0.2cm}56752.187  & \hspace{-0.3cm} 11.33	$\pm$0.07 	   & \hspace{-0.3cm}10.90$\pm$0.07   & \hspace{-0.3cm}10.65$\pm$0.07	  \\
  \hline
\end{tabular}\label{IR_phot}  
\end{table*}

\end{document}